\documentclass[sigconf]{acmart}
\usepackage{multirow}
\AtBeginDocument{%
  }

\copyrightyear{2025} 
\acmYear{2025} 
\setcopyright{cc}
\setcctype{by}
\acmConference[MM '25]{Proceedings of the 33rd ACM International Conference on Multimedia}{October 27--31, 2025}{Dublin, Ireland}
\acmBooktitle{Proceedings of the 33rd ACM International Conference on Multimedia (MM '25), October 27--31, 2025, Dublin, Ireland}\acmDOI{10.1145/3746027.3755814}
\acmISBN{979-8-4007-2035-2/2025/10}

\begin{document}

\title{DT-UFC: Universal Large Model Feature Coding via Peaky-to-Balanced Distribution Transformation}

\author{Changsheng Gao}
\email{changsheng.gao@ntu.edu.sg}
\affiliation{%
  \institution{Nanyang Technological University}
  \city{Singapore}
  \country{Singapore}
}
\author{Zijie Liu}
\email{jeffrey101@stu.xmu.edu.cn}
\affiliation{%
  \institution{Xiamen University}
  \city{Xiamen}
  \country{China}
}
\author{Li Li}
\email{lil1@ustc.edu.cn}
\affiliation{%
  \institution{University of Science and Technology of China}
  \city{Hefei}
  \country{China}
}
\author{Dong Liu}
\email{dongeliu@ustc.edu.cn}
\affiliation{%
  \institution{University of Science and Technology of China}
  \city{Hefei}
  \country{China}
}
\author{Xiaoyan Sun}
\email{sunxiaoyan@ustc.edu.cn}
\affiliation{%
  \institution{University of Science and Technology of China}
  \city{Hefei}
  \country{China}
}
\author{Weisi Lin\textsuperscript{\textdagger}}
\email{wslin@ntu.edu.sg}
\affiliation{%
  \institution{Nanyang Technological University}
  \city{Singapore}
  \country{Singapore}
}

\renewcommand{\shortauthors}{Changsheng Gao et al.} 

\begin{abstract}
Like image coding in visual data transmission, feature coding is essential for the distributed deployment of large models by significantly reducing transmission and storage burden. However, prior studies have mostly targeted task- or model-specific scenarios, leaving the challenge of universal feature coding across diverse large models largely unexplored. In this paper, we present the first systematic study on universal feature coding for large models. The key challenge lies in the inherently diverse and distributionally incompatible nature of features extracted from different models. For example, features from DINOv2 exhibit highly peaky, concentrated distributions, while those from Stable Diffusion 3 (SD3) are more dispersed and uniform. This distributional heterogeneity severely hampers both compression efficiency and cross-model generalization. To address this, we propose a learned peaky-to-balanced distribution transformation, which reshapes highly skewed feature distributions into a common, balanced target space. This transformation is non-uniform, data-driven, and plug-and-play, enabling effective alignment of heterogeneous distributions without modifying downstream codecs. With this alignment, a universal codec trained on the balanced target distribution can effectively generalize to features from different models and tasks. We validate our approach on three representative large models (LLaMA3, DINOv2, and SD3) across multiple tasks and modalities. Extensive experiments show that our method achieves notable improvements in both compression efficiency and cross-model generalization over task-specific baselines. All source code has been made available at \url{https://github.com/chansongoal/DT-UFC}.
\end{abstract}




\maketitle

\begin{figure}[htp]
    \centering
    \includegraphics[width=0.95\columnwidth]{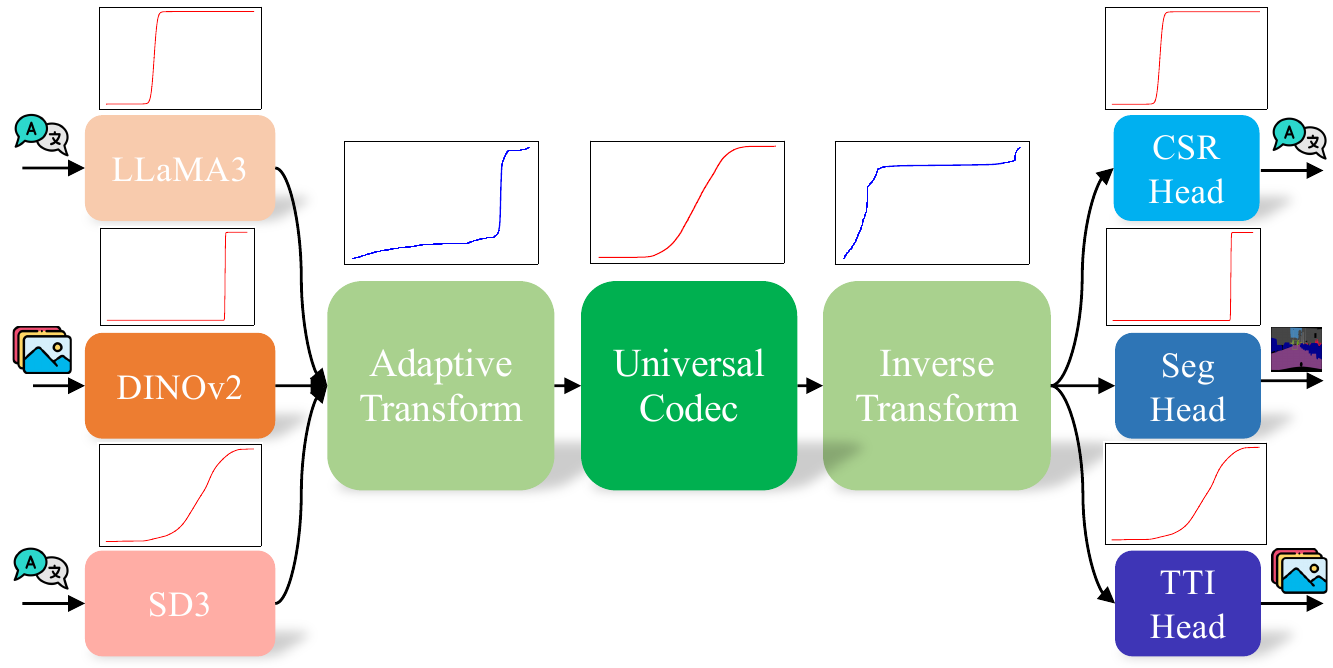}
    \caption{The overall framework of DT-UFC. Large models generate heterogeneous features (red CDF curves). The proposed adaptive transforms (blue curves) reshape these features into a unified target distribution space. The decoded features are inverse-transformed and passed to task-specific heads to complete predictions. (Sec. \ref{subsec_framework})}
    \label{fig_framework}
\end{figure}
\section{Introduction}
In recent years, large foundation models such as GTP-4 \cite{achiam2023gpt}, DeepSeek \cite{guo2025deepseek}, and Gemini \cite{team2024gemini} have demonstrated remarkable capabilities across a wide range of reasoning, discriminative, and generative tasks. As these models are increasingly integrated into distributed and resource-constrained systems \cite{tian2022fedbert, Sagawa2020Distributionally,ye2024openfedllm}, the transmission of intermediate representations has become a bottleneck for scalable deployment. Similar to image and video coding \cite{li2024ustctd,gao2022twostep,zuo2024learned,mao2024perceptual,li2024object,tang2024offline,li2024inloop,li2022global,li2024uniformly}, feature coding aims to encode intermediate features compactly while preserving semantic fidelity. 
However, existing feature coding approaches are predominantly model- or task-specific \cite{henzel2022efficient, wang2024low, ma2024feature, alvar2019multi, yan2021SSSIC,chen2024end}, where the encoder is tailored to a particular feature distribution. Such specialization severely limits their universality in realistic scenarios where multiple large models coexist and are applied to various downstream tasks.

These limitations highlight the need for \textbf{universal feature coding} — a single codec capable of compressing and reconstructing intermediate representations from diverse models and tasks. Despite its practical importance, universal feature coding remains largely unexplored. The fundamental challenge lies in the distributional heterogeneity of large model features.
As illustrated in Fig. 2, features extracted from different models exhibit diverse statistical properties and value distributions. For example, features from DINOv2 \cite{oquab2023dinov2} exhibit highly peaky distributions with dense value concentrations, while Stable Diffusion 3 (SD3) \cite{esser2024sd3} features are more evenly spread over a narrower range. Such heterogeneity stems from fundamental differences in model architecture, training objectives, and source modalities. This poses a severe obstacle to universal feature coding: a codec trained on a specific feature distribution often generalizes poorly to others. 
Moreover, these peaky distributions introduce additional challenges for feature coding: they often lead to sparse latent activations, poor entropy modeling efficiency, and unstable codec training. These issues jointly degrade compression performance and further hinder generalization to unseen models or tasks.

To tackle these challenges, we propose a simple yet reasonable insight: rather than forcing a universal encoder to directly accommodate heterogeneous inputs, we introduce a learnable transformation module that reshapes peaky, model-specific feature distributions into a common, balanced target space. By aligning feature distributions through a preprocessing step, we decouple the codec from model-specific statistics, enabling robust generalization across feature distributions. This idea draws inspiration from classical signal processing, where histogram equalization improves encoding robustness by balancing value distributions.

Building on this insight, we propose \textbf{DT-UFC}, a novel framework for \textbf{U}niversal \textbf{F}eature \textbf{C}oding via Peaky-to-Balanced \textbf{D}istribution \textbf{T}ransformation. The core of DT-UFC is the learnable non-uniform peaky-to-balanced transform that can adaptively transforms diverse input feature distributions into a common, balanced target distribution.
This transformation is designed to be plug-and-play and does not require any modification to the downstream codec.
Once transformed, the features can be compressed by a universal codec and then inversely transformed for downstream tasks. This not only bridges the distribution gap between features but also enhances the expressiveness of the latent representation and improves entropy coding efficiency.
To evaluate the generalization ability of DT-UFC, we conduct experiments across three representative tasks: common sense reasoning (CSR), semantic segmentation (Seg), and text-to-image generation (TTI).
These tasks cover both discriminative and generative paradigms and span vision and language modalities, providing a comprehensive evaluation for universal feature coding.
Our main contributions are summarized as follows:
\begin{itemize}
    \item We formally introduce the problem of universal feature coding and identify distributional heterogeneity as the key bottleneck preventing generalization across various models and tasks.
    \item We propose DT-UFC, a plug-and-play framework that learns a peaky-to-balanced transformation to align heterogeneous feature distributions into a balanced target space, enabling a single codec to handle diverse inputs effectively.
    \item We demonstrate the effectiveness and generalization ability of DT-UFC across multiple large models, tasks, and modalities, achieving superior rate-distortion performance compared to task-specific baselines.
\end{itemize}

\begin{figure*}[h]
    \centering
    \setlength{\tabcolsep}{2pt}
    \resizebox{0.9\textwidth}{!}{%
    \begin{tabular}{cccc}
        \includegraphics[height=4cm]{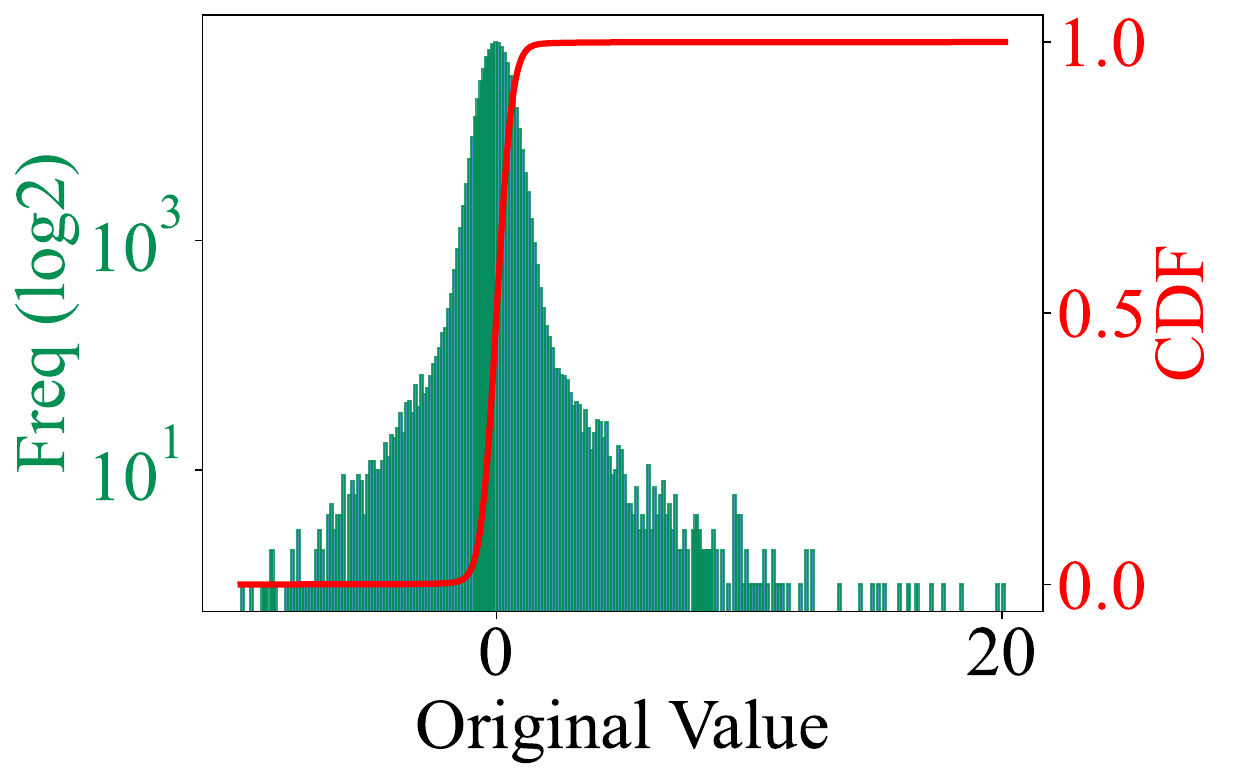} &
        \includegraphics[height=4cm]{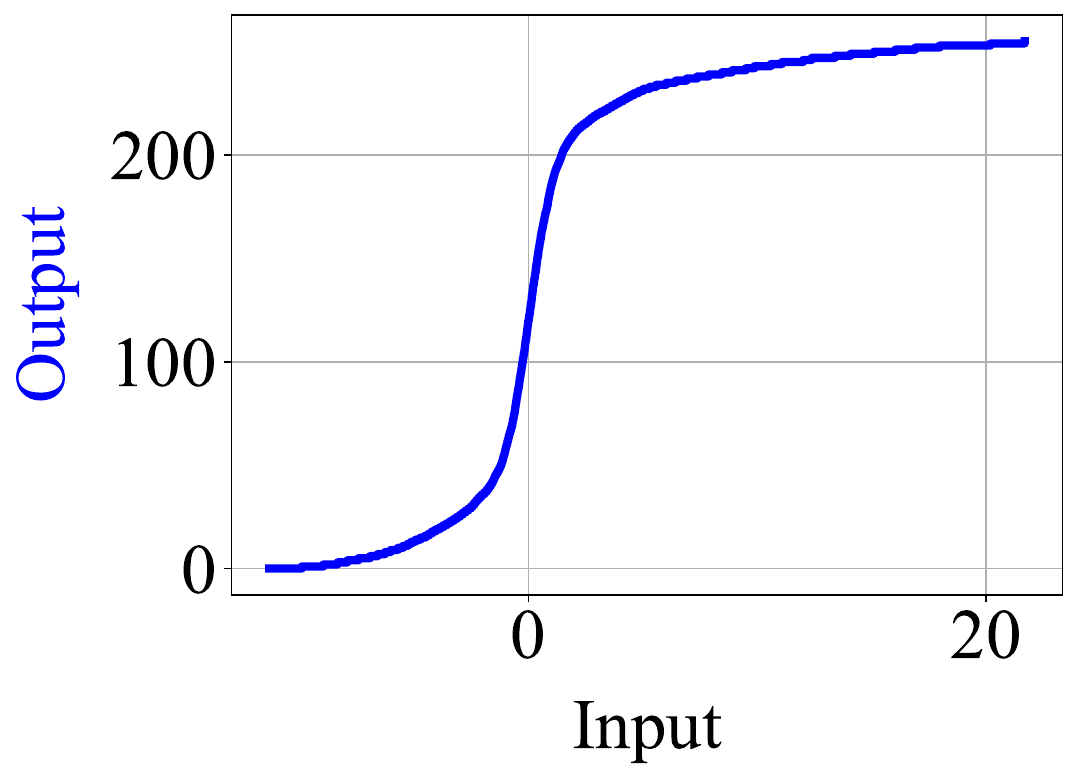} &
        \includegraphics[height=4cm]{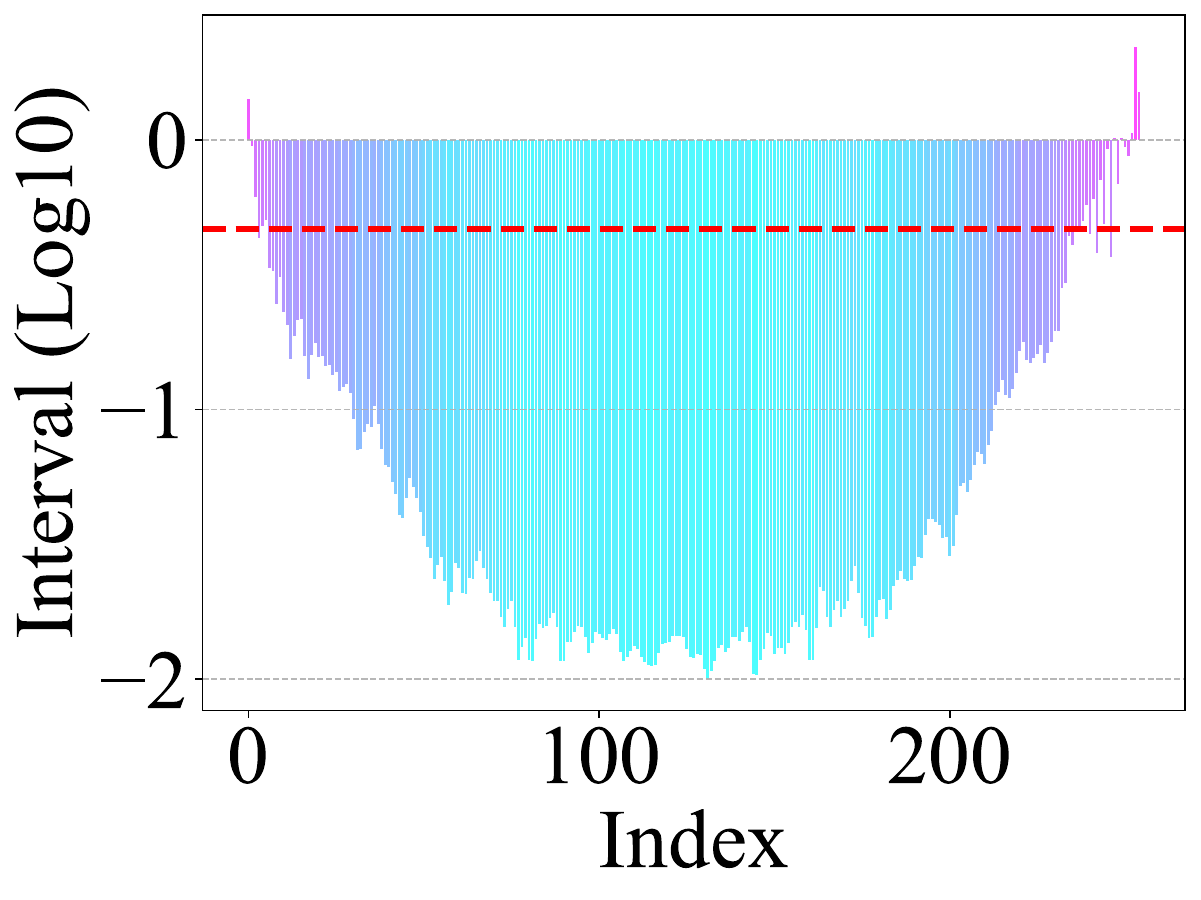} &
        \includegraphics[height=4cm]{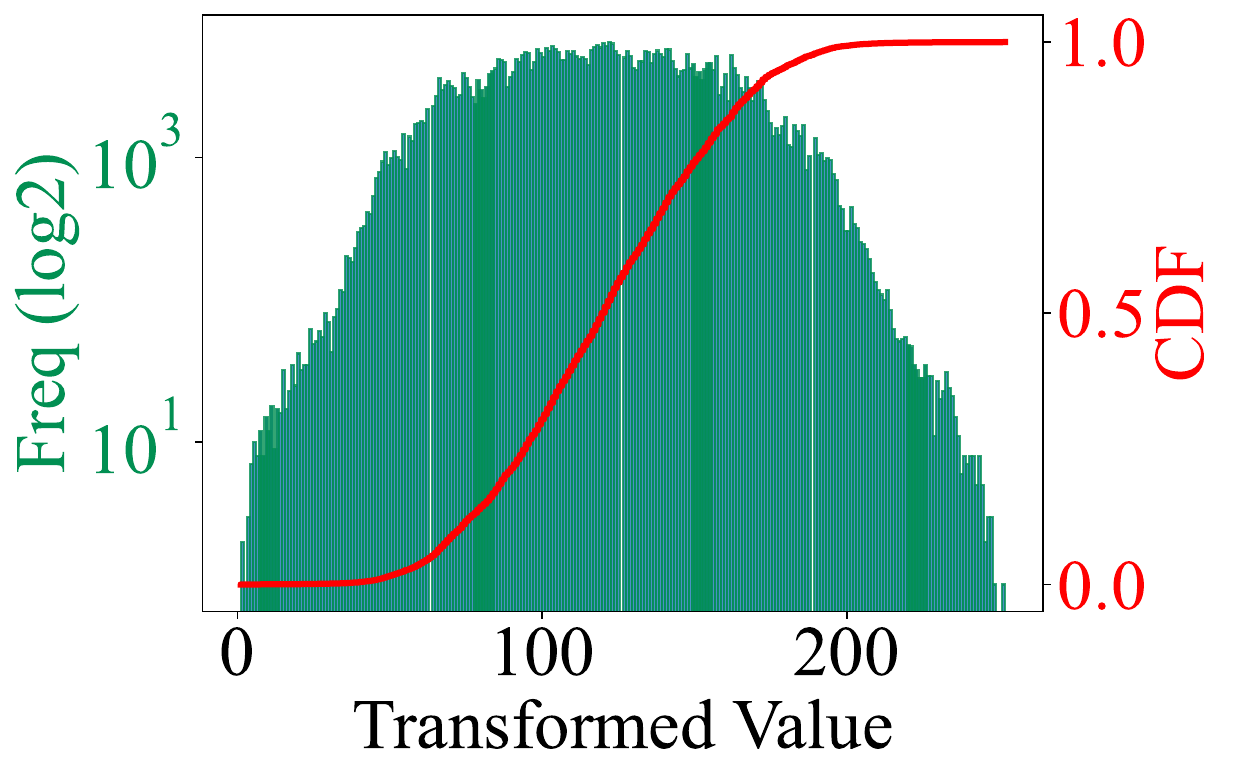} \\

        \includegraphics[height=4cm]{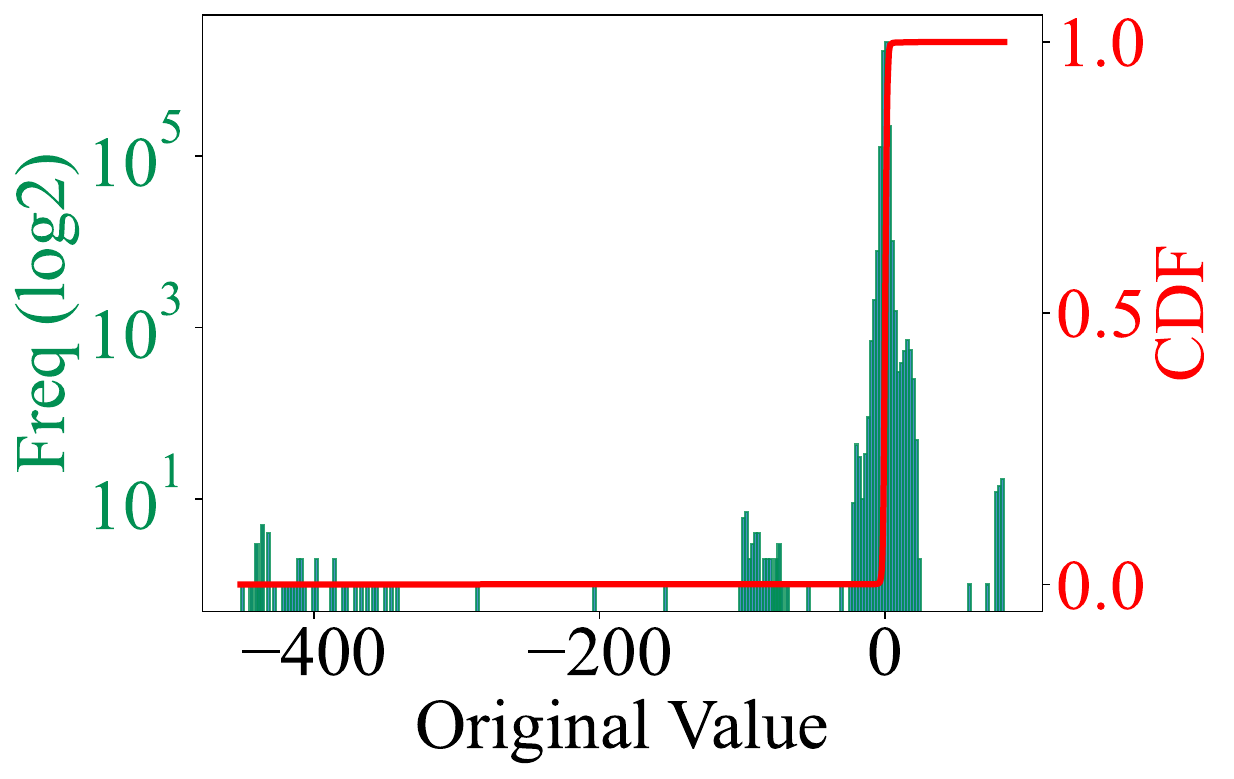} &
        \includegraphics[height=4cm]{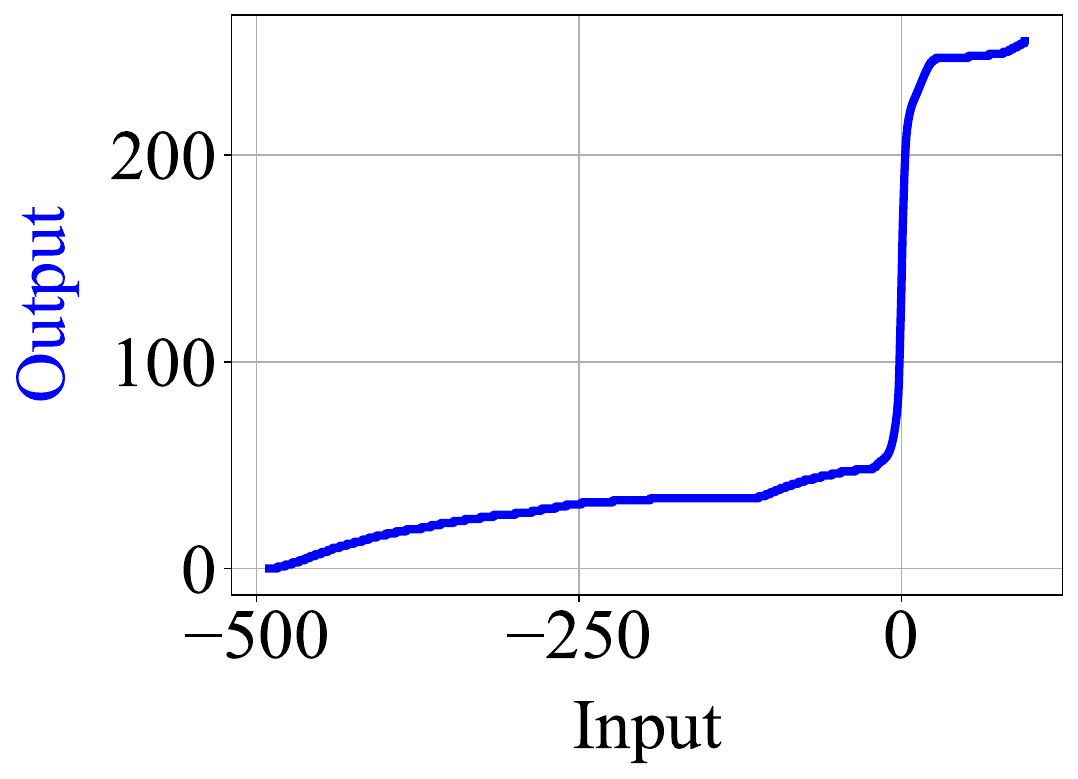} &
        \includegraphics[height=4cm]{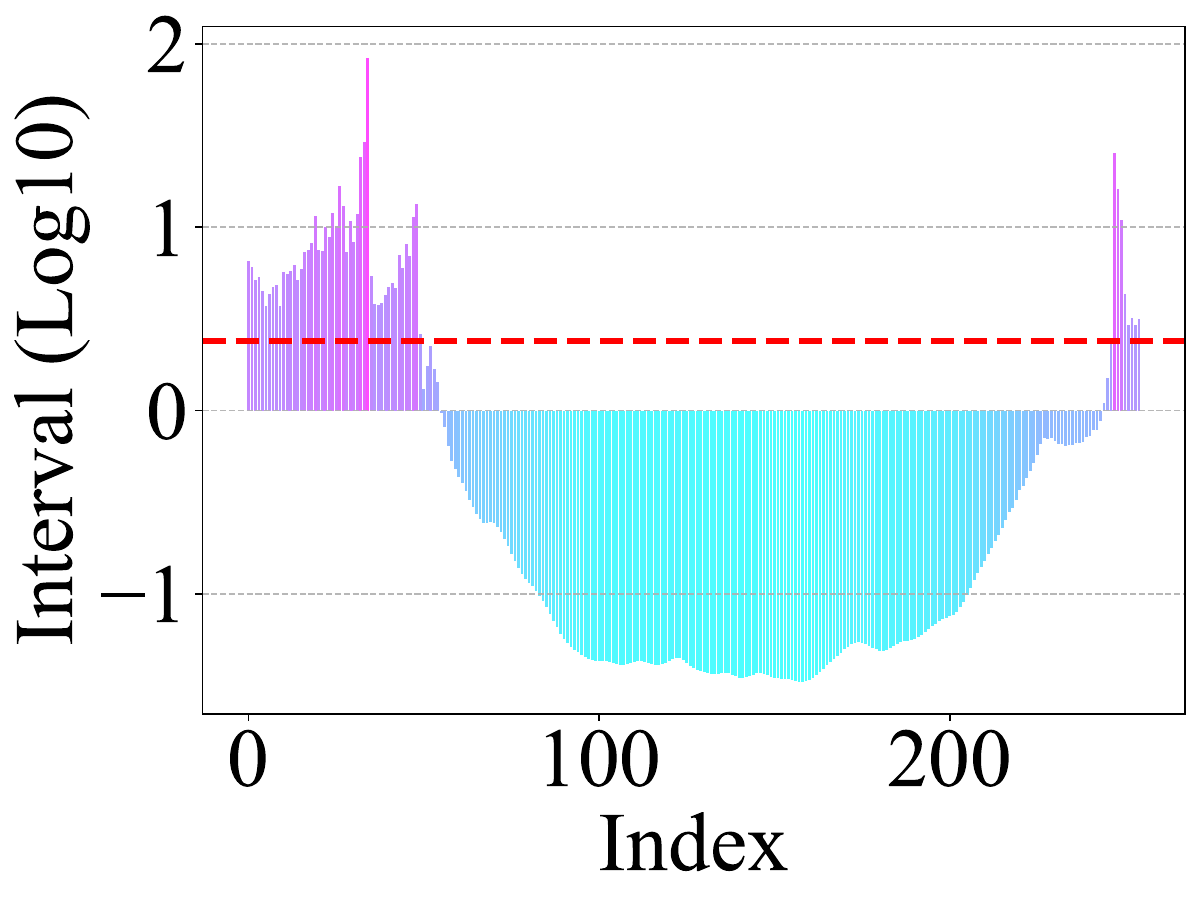} &
        \includegraphics[height=4cm]{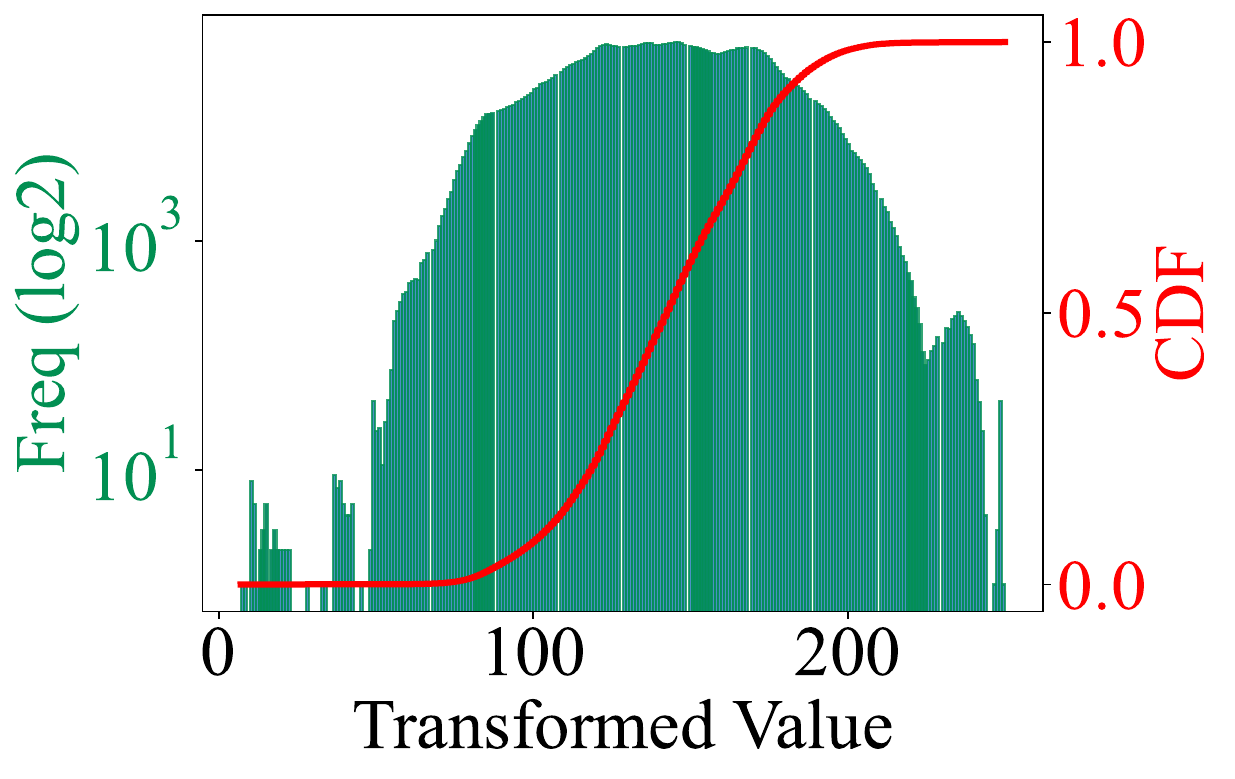} \\

        \includegraphics[height=4cm]{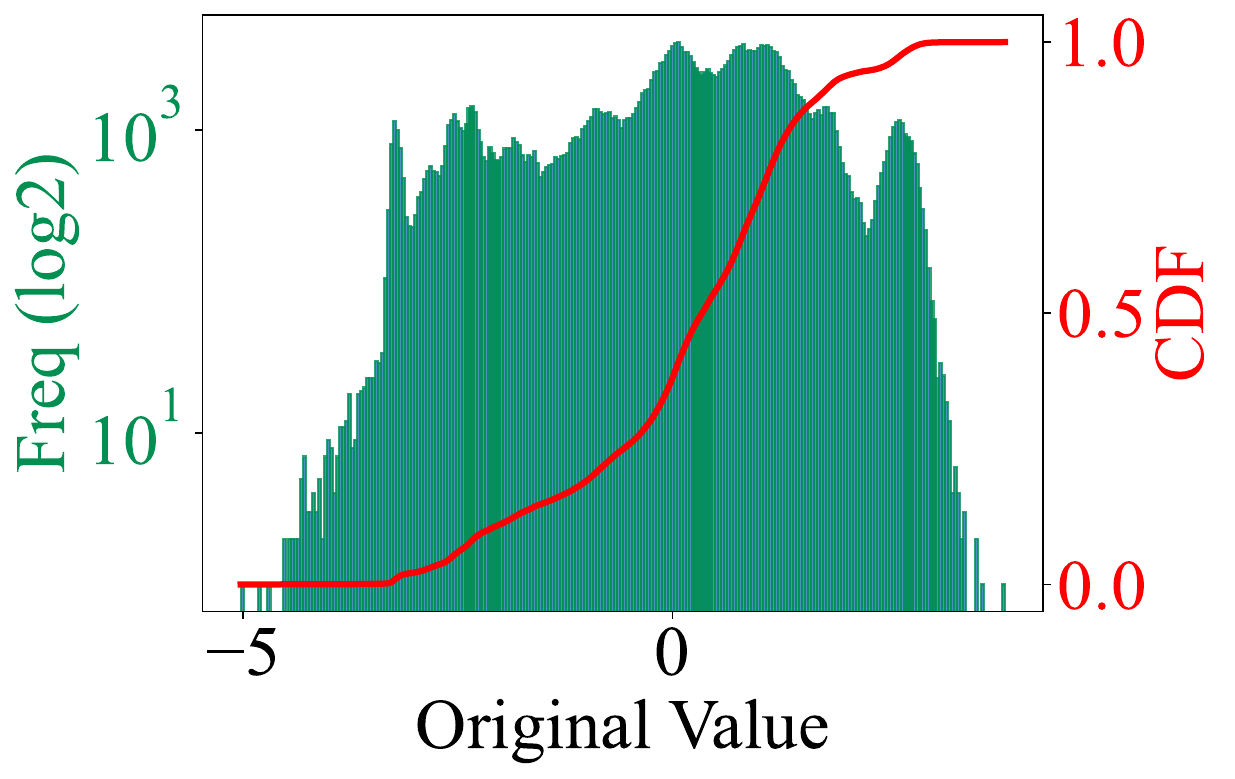} &
        \includegraphics[height=4cm]{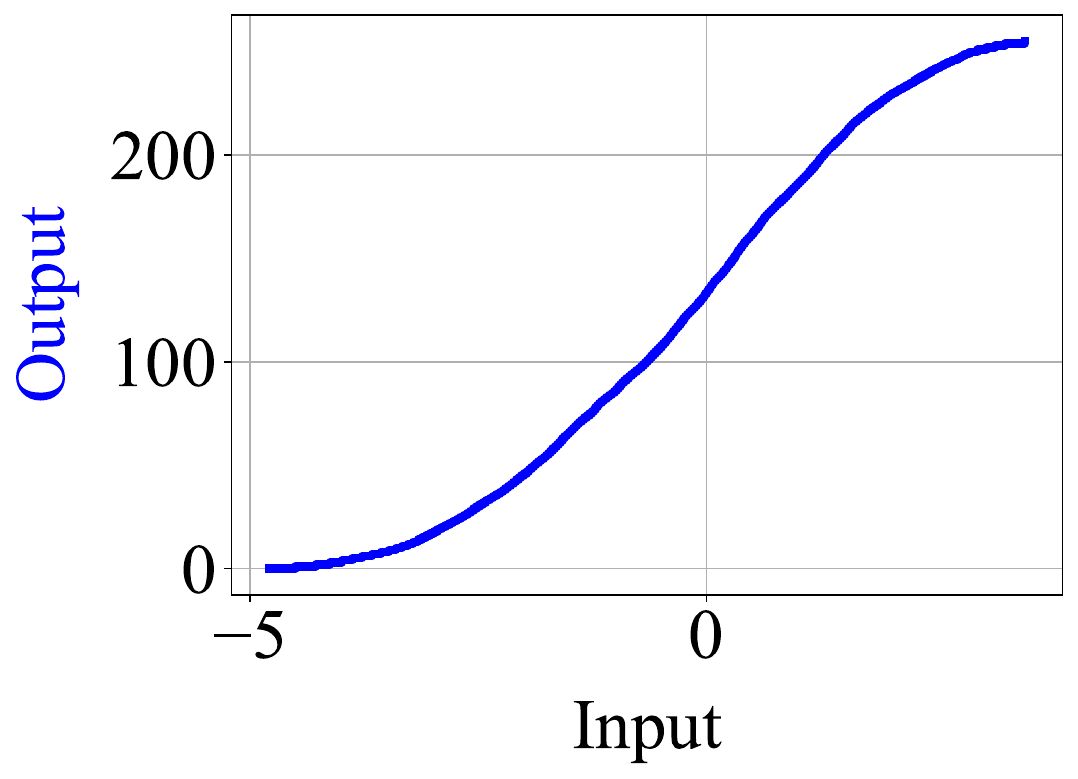} &
        \includegraphics[height=4cm]{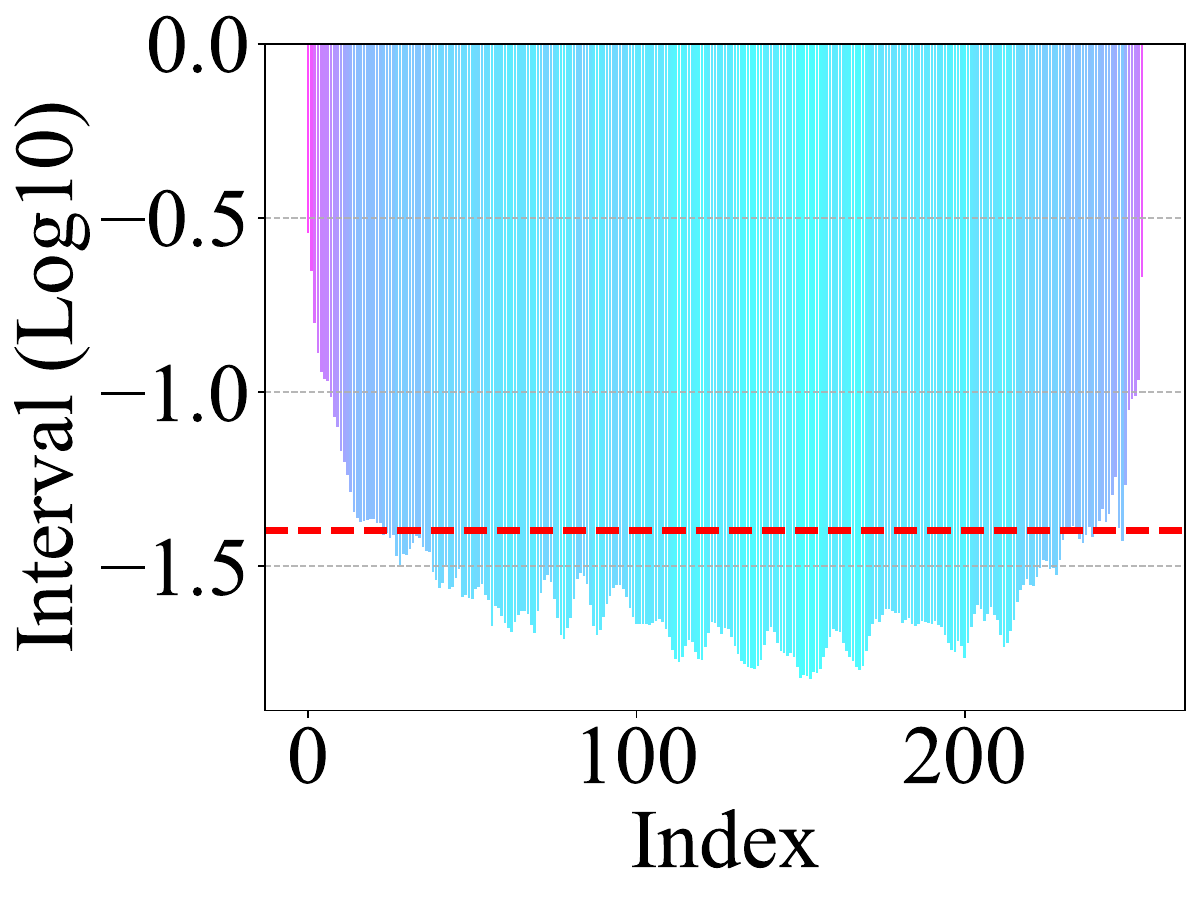} &
        \includegraphics[height=4cm]{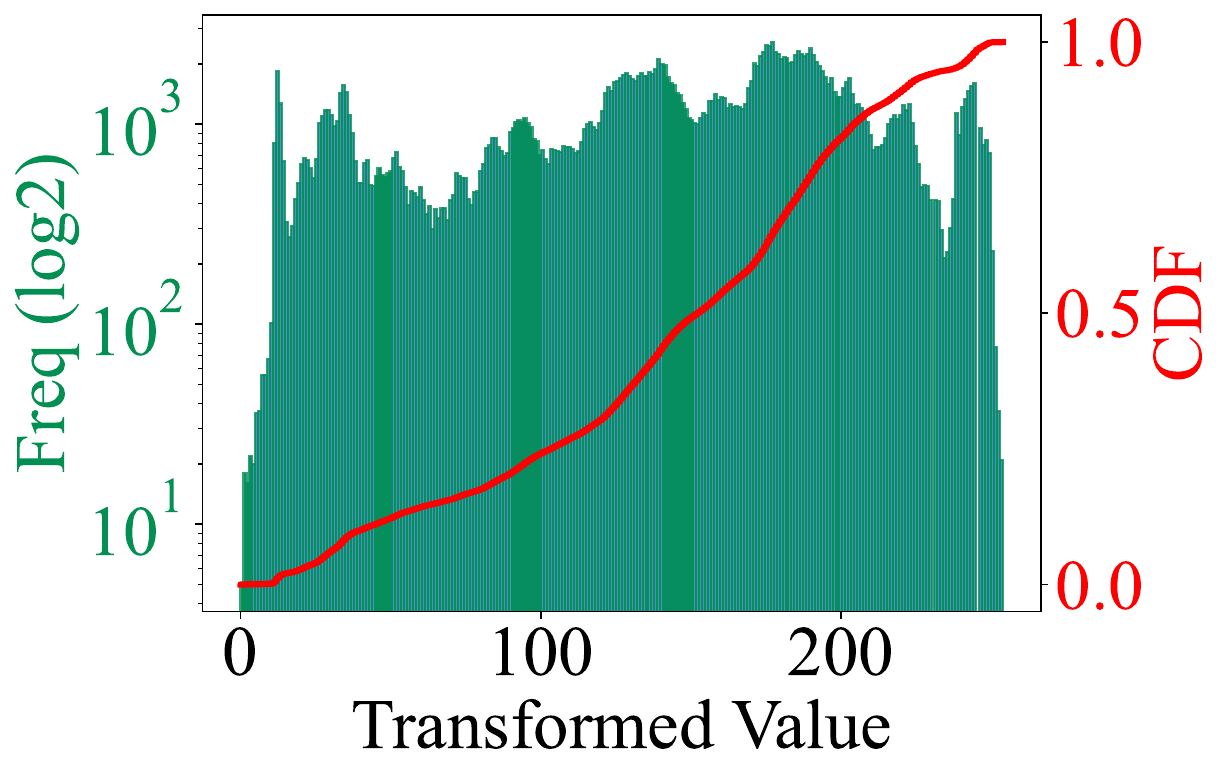}
    \end{tabular}
    }
    \caption{Visualizations of the original feature distribution (1st column), the nonlinear transformation function (2nd column), the width of the split intervals in the original feature space (3rd column), and the transformed feature distribution (4th column). The first, second, and third rows correspond to the CSR, Seg, and TTI tasks, respectively. (Sec. \ref{subsec_transform_derivation})}
    \label{fig_distribution}
\end{figure*}

\section{Related Work}
\label{sec_related_work}
\subsection{Feature Coding}
Feature coding, as a key branch of coding for machines \cite{gao2023towards,zhu2024learned,yin2025unified,shen2024image,shen2024image2,liu2024rate,chen2023transtic,kao2024bridging,tian2023non,tian2024free,tian2024coding}, was originally introduced to compress intermediate features in collaborative intelligence scenarios, where deep models are split across edge and cloud. Early work in \cite{chen2019lossy} evaluated coding performance using standard image codecs but suffered from low efficiency due to distribution mismatch. Subsequent research focused on task-specific feature coding \cite{choi2018deep,cai2022high,henzel2022efficient,wang2024low,ma2024feature,gao2024rethinking}, where compression pipelines are tailored to single-task scenarios like image classification or detection. While effective, these methods lack generalizability across diverse tasks or models.

To broaden applicability, recent efforts explored multi-task feature coding \cite{alvar2019multi,yan2021SSSIC,alvar2020bit,feng2022image,gao2024dmofc,gao2024imofc,wang2022towards,zhang2021MSFC,guo2023toward,wang2023intermediate}. For instance, \cite{yan2021SSSIC} proposed a scalable coding scheme for coarse-to-fine tasks. However, these methods typically rely on task-specific decoders or side information and remain constrained to fixed sets of tasks or models. Recognizing the growing importance of feature coding, standardization efforts have also emerged, notably through MPEG’s ongoing work on feature coding standards \cite{fcm_cfp}. 

Both academic research and standardization have predominantly focused on features extracted from CNN-based models \cite{zhang2021MSFC,kim2023end,misra2022video,liu2021semantics,wu2025codebook}. 
In contrast, the current landscape of artificial intelligence has been dominated by large models \cite{guo2025deepseek,achiam2023gpt,bai2023qwen,oquab2023dinov2,touvron2023llama, esser2024sd3}. 
To bridge this gap, \cite{gao2024feature} introduced the first large model feature coding benchmark, providing a dataset, unified test conditions, and baseline methods. This work highlights the critical role of feature coding in large model deployment. However, the proposed baselines require model-specific training and do not address the universal feature coding problem.

Therefore, a universal feature codec that can handle diverse models and tasks is increasingly needed. 
Our work takes a step toward this goal by proposing a learnable transformation that aligns feature distributions, enabling efficient, model-agnostic compression with a single codec.

\subsection{Feature Alignment}
Feature alignment is a common technique in representation learning to reduce domain shift and improve generalization. It has been widely applied in domain adaptation \cite{joonho2023feature, yeh2021sofa,long2015learning, sun2016deep,wang2023feature}, multi-modal learning \cite{radford2021learning,li2021align,li2022blip,li2023blip,ge2021structured}, and multi-task learning \cite{wang2022domain,senushkin2023independent}, enabling models to share or transfer representations across distributions, modalities, or tasks.

Despite its effectiveness, feature alignment has never been applied to feature coding. Existing codecs typically assume fixed input statistics and fail to generalize across heterogeneous features. In this work, we introduce alignment as a pre-coding transformation, enabling a universal codec to handle diverse large model features for the first time.

\section{Challenges in Universal Feature Coding}
\label{sec_challenge}
This section outlines the key challenges hindering universal feature coding across large models.
\subsection{Feature Distribution Analysis}
A core challenge of universal feature coding lies in the heterogeneous nature of feature distributions across models, stemming from differences in architecture, task objectives, and training data.
We visualize the feature distribution in Fig. \ref{fig_distribution}.
The green histogram shows the log-scaled frequency of the original feature values, while the red curve represents their empirical cumulative distribution function (CDF). 

The feature distributions vary significantly in both statistical range and shape.
DINOv2 features exhibit a wide value range with a majority of values densely concentrated in a narrow interval, resulting in a highly peaky distribution. LLaMA3 features follow a similar trend, though with a slightly narrower range and less extreme concentration. In contrast, SD3 produces features that are more balanced and flatly distributed over a limited range. 
These differences highlight the incompatibility of feature distributions across models, posing a fundamental challenge to universal coding.

\begin{table}[tp]
    \caption{KL Divergence Comparison between the Original and Transformed Feature Distributions (Sec. \ref{subsec_challenge_universality})}
    \label{tab_kl}
    \resizebox{0.3\textwidth}{!}{%
    \begin{tabular}{@{}cccc@{}}
    \toprule
    \textbf{Original}    & \textbf{CSR ($q$)} & \textbf{Seg ($q$)} & \textbf{TTI ($q$)} \\ \midrule
    \textbf{CSR ($p$)}  & 0            & 19.87        & 2.95         \\
    \textbf{Seg ($p$)}  & 18.06        & 0            & 4.92         \\
    \textbf{TTI ($p$)}  & 2.95         & 15.18        & 0            \\ \midrule \midrule
    \textbf{Transformed} & \textbf{CSR ($q$)} & \textbf{Seg ($q$)} & \textbf{TTI ($q$)} \\ \midrule
    \textbf{CSR ($p$)}  & 0            & 0.50         & 0.57         \\
    \textbf{Seg ($p$)}  & 0.25         & 0            & 0.50         \\
    \textbf{TTI ($p$)}  & 1.15         & 2.23         & 0           \\ \bottomrule
    \end{tabular}}
\end{table}
\subsection{Challenges in Universality}
\label{subsec_challenge_universality}
Encoders trained on features from a specific large model tend to internalize distribution-specific priors and inductive biases. 
When applied to features from another model with a substantially different statistical distribution, these learned priors are violated, resulting in a significant domain shift that degrades both reconstruction quality and generalization performance \cite{Sagawa2020Distributionally}.

Moreover, peaky feature distributions further exacerbate the generalization issue. Such inputs tend to encourage neural networks to memorize dominant patterns while neglecting rare but semantically meaningful variations \cite{zhang2021understanding, arpit2017closer}. This tendency reduces the network’s ability to learn robust representations, especially when exposed to data outside the training domain.
As a result, the learned representations become brittle and narrowly focused, unable to generalize across diverse input variations or support cross-distribution deployment.

To validate this, we characterize the similarity between different feature distributions using Kullback-Leibler (KL) divergence. Each feature distribution is approximated using a discretized histogram. The KL divergence is computed as 
$D_{\mathrm{KL}}(p \| q) = \sum p \log (p/q)$.
As shown in Table~\ref{tab_kl}, original features from different models exhibit large KL divergence values, confirming their inherent distributional misalignment. This implies that an encoder trained on one feature distribution must allocate additional bitrates to accurately compress features from another distribution.
In contrast, the KL divergence among transformed feature distributions is significantly lower, demonstrating that our proposed distribution transformation (see Section~\ref{sec_method}) effectively aligns statistical properties across different models. This alignment serves as a critical enabler for universal feature coding.

\subsection{Challenges in Coding Performance}
\label{subsec_challenge_performance}
Beyond generalization, peaky input distributions also negatively impact the efficiency and expressiveness of learned encoders. Specifically, they restrict the effective exploitation of the latent space, destabilize quantization, and reduce entropy coding efficiency.

\begin{figure}[tp]
    \centering
    \includegraphics[width=0.48\columnwidth]{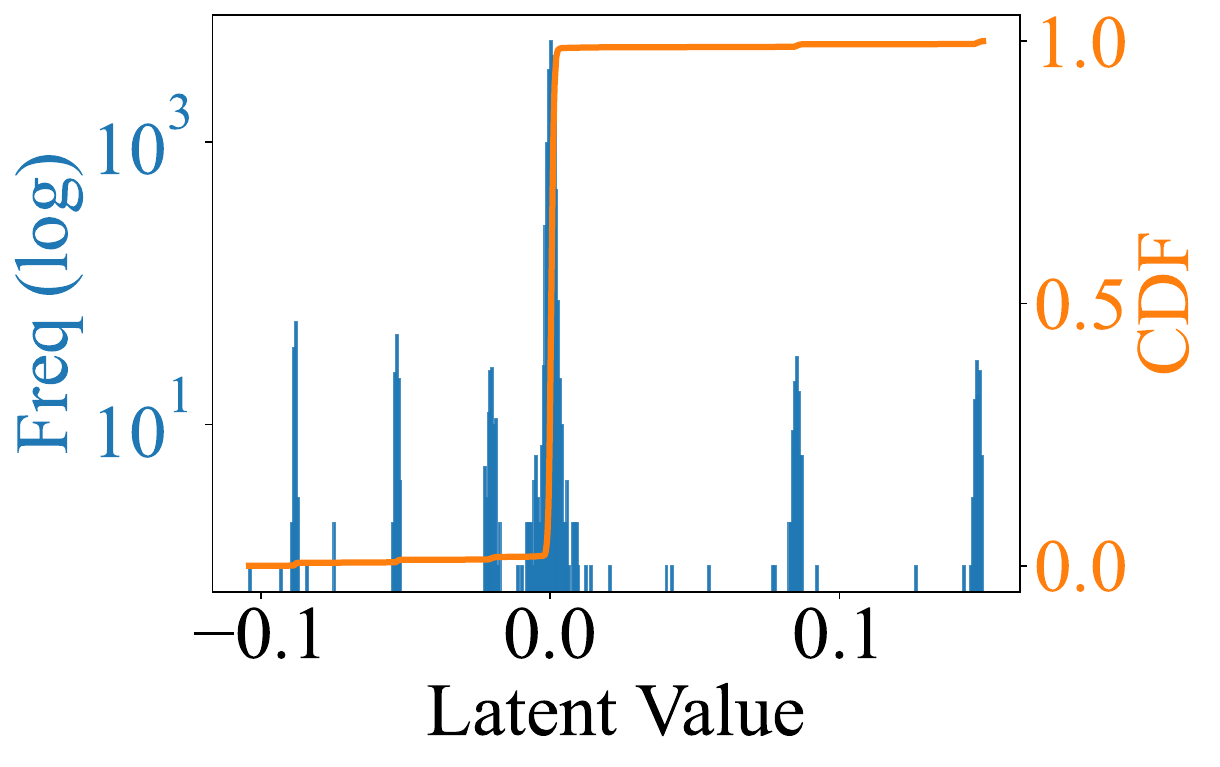}
    \includegraphics[width=0.48\columnwidth]{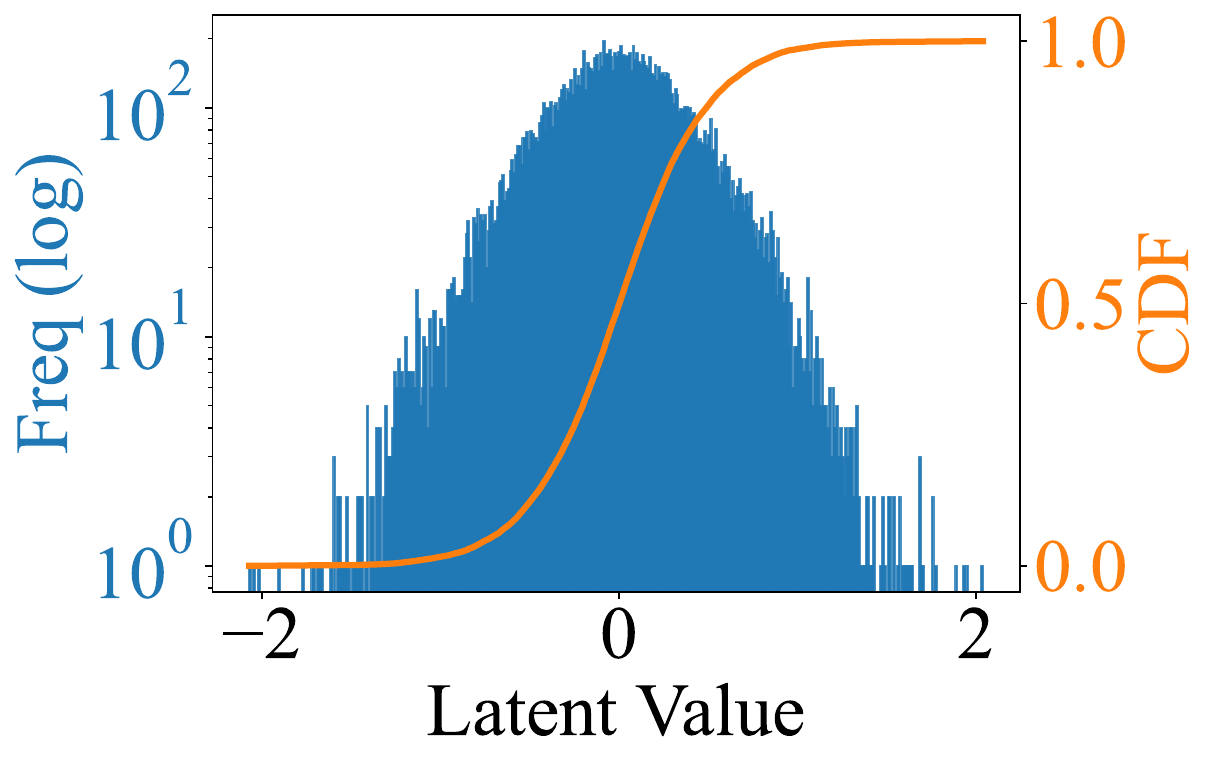}
    \caption{Visualization of the latent space. Left: latent representations obtained from an encoder trained on the original Seg features. Right: latent representations obtained from an encoder trained on the transformed Seg features. (Sec. \ref{subsec_challenge_performance})}
    \label{fig_latent}
\end{figure}

When most of the input mass resides in a narrow range, the encoder’s analysis transform is biased toward over-representing these dominant intervals. Consequently, the latent space is sparsely and unevenly activated, as shown in Fig.~\ref{fig_latent}, where only a few regions are heavily populated while large portions remain underutilized. 
Such non-uniform activation patterns limit the encoder’s representational capacity and are suboptimal for downstream entropy modeling. In addition, this indicates that the encoder transform capacity is not fully exploited, leading to limited transform capacity and poor adaptability to out-of-distribution inputs.

This sparse activation pattern violates the Lloyd-Max principle, which dictates that optimal quantizers should allocate bins proportional to the input density to minimize quantization distortion. However, under peaky distributions, large parts of the quantization space are wasted, resulting in inefficient bit allocation and increased distortion.

In addition, this non-uniform latent usage undermines the assumptions underlying many entropy models, which typically rely on smooth and continuous latent distributions for accurate likelihood estimation. Sharp discontinuities and empty regions lead to unstable training, inaccurate bitrate estimation, and ultimately deteriorate rate–distortion performance.

\section{The Proposed DT-UFC Method}
\label{sec_method}

\subsection{Overall Framework}
\label{subsec_framework}
Based on the challenges discussed in Section~\ref{sec_challenge}, we identify a critical requirement for universal feature coding:
\begin{center}
    \textbf{Feature distributions should be aligned across models and avoid excessive concentration.}
\end{center}

This insight motivates us to seek a feature transformation that reshapes diverse feature distributions into a common and balanced target distribution. Such a transformation reduces distributional discrepancies and enables a universal encoder to compress heterogeneous features efficiently. 

We propose the DT-UFC framework, illustrated in Fig.~\ref{fig_framework}. Given that input features are extracted from various large models, we first apply an adaptive transformation that normalizes and redistributes the original feature values into a common, balanced distribution space. A universal encoder is then trained to compress features in this space. After decoding, the features are mapped back to their original domain via an inverse transformation, and then fed into task-specific heads for inference.

The core challenge in the DT-UFC framework lies in designing a transformation that aligns heterogeneous input distributions to a common target, while minimizing transformation-induced distortion.
\begin{figure*}[h]
  \centering
    \resizebox{0.9\textwidth}{!}{%
    \begin{tabular}{ccc}
        \includegraphics[height=4cm]{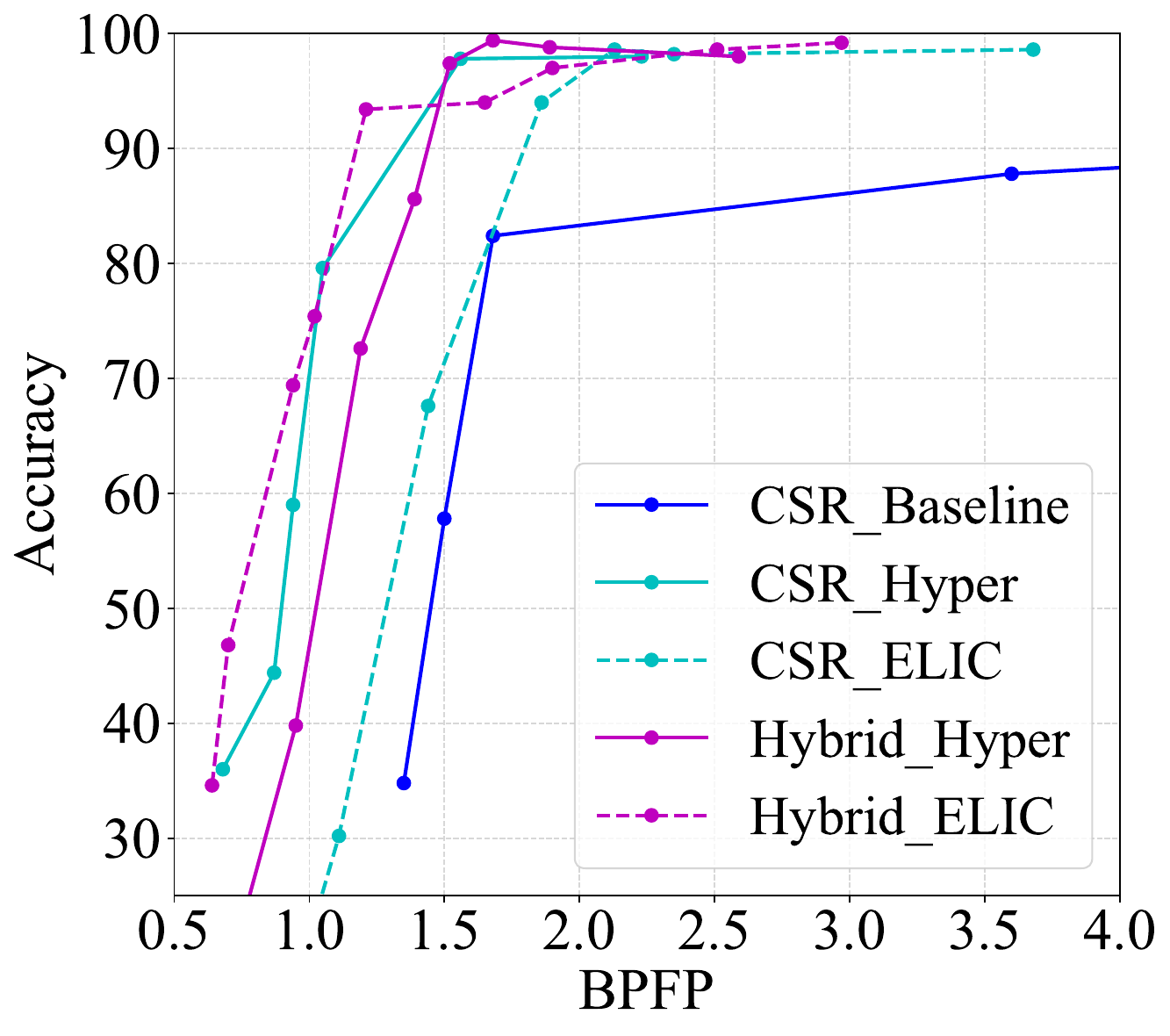} &
        \includegraphics[height=4cm]{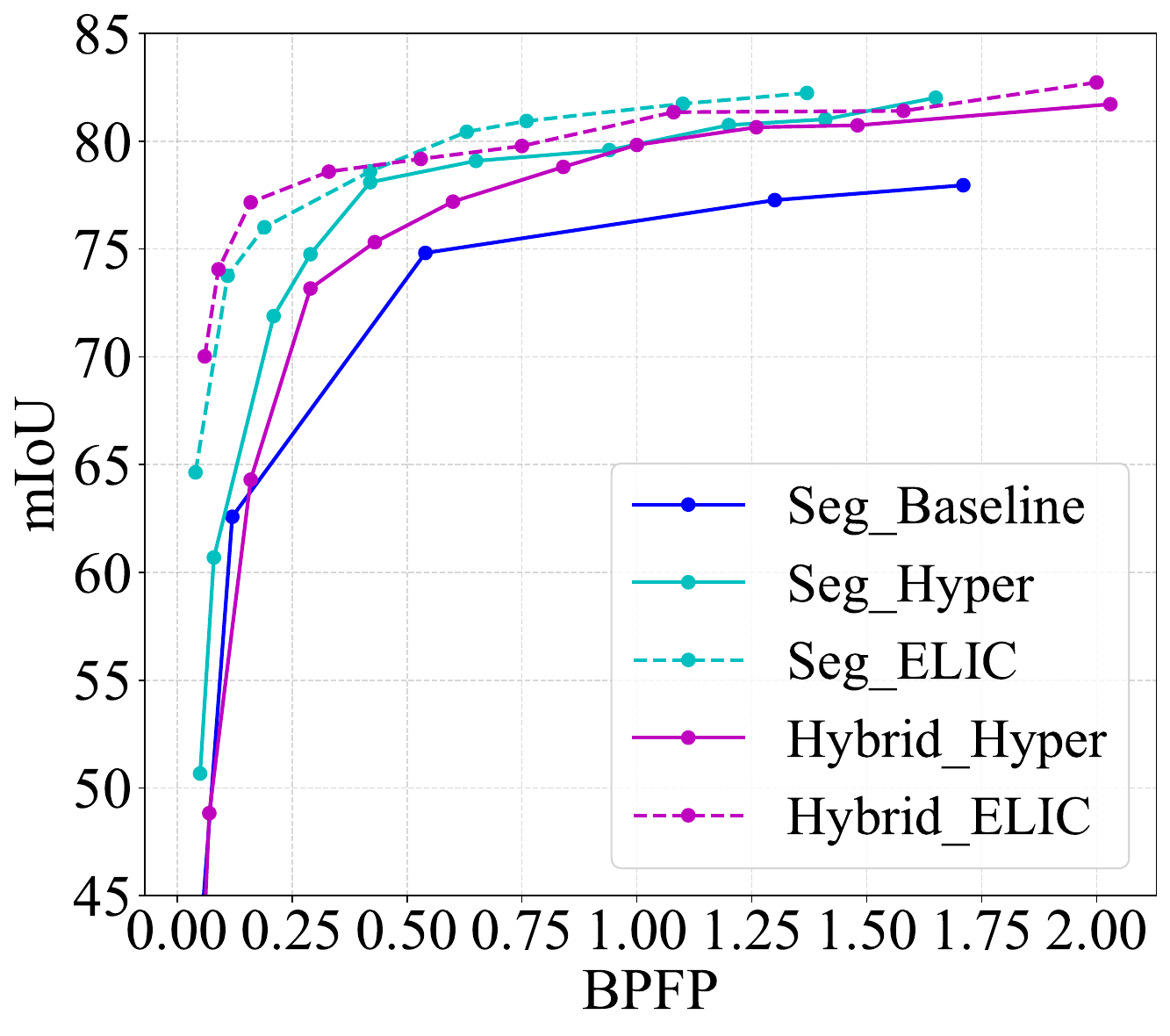} &
        \includegraphics[height=3.95cm]{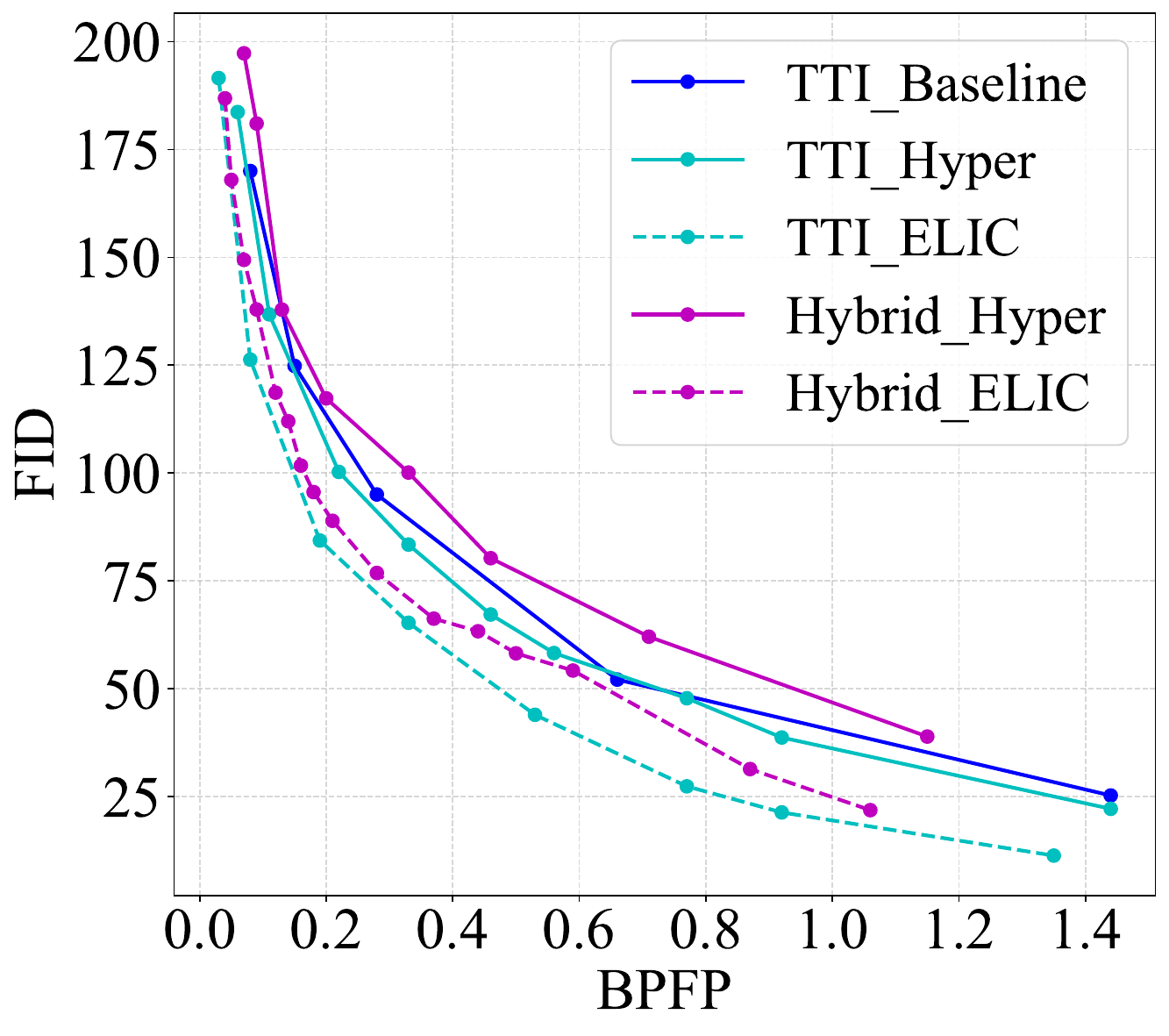} \\
    \end{tabular}
    }
  \caption{Rate-accuracy performance comparisons between the baseline and the proposed method. The left, middle, and right figures correspond to the CSR, Seg, and TTI tasks, respectively. (Sec. \ref{subsec_rd_performance})}
  \label{fig_rd}
\end{figure*}
\subsection{Non-uniform Feature Transformation}
\label{subsec_transform_derivation}
To address the problem outlined above, we first define two essential criteria for a desirable transformed feature distribution:
\begin{itemize}
    \item \textbf{Consistent Distribution Range:} Feature values from different models should be mapped to a common range to facilitate model-agnostic feature coding.
    \item \textbf{Balanced Distribution:} Feature values should be evenly distributed within the range to fully exploit the latent capacity of the encoder and enable efficient entropy coding.
\end{itemize}

To satisfy the first criterion, we decided to map the original features into a fixed space. Inspired by image coding, we define a fixed target space of $p_t \in \{0, 1, \dots, 255\}$ for transformed features. This standardization facilitates compatibility with existing image codecs and allows for direct reuse of quantization and entropy modeling techniques.
To satisfy the second criterion, we design a learnable non-uniform transformation function that redistributes the original feature values to form a balanced histogram in the target space. This is especially important for features from models like LLaMA3 and DINOv2, whose raw distributions are highly concentrated.

Instead of deriving a closed-form function, we treat the transformation as a data-driven optimization problem. We aim to partition the input space into 256 non-overlapping regions and map each region to a unique integer in $p_t$. Now the problem turns to how to split the original distribution range into 256 regions with minimal transformation distortion.

Formally, given an input feature $\mathbf{f} \in \mathbb{R}^d$ sampled from distribution $p_f$, we define a transformation $T: \mathbb{R} \rightarrow \{0, 1, \dots, 255\}$ such that:
\begin{itemize}
    \item The range of $\mathbf{f}$ is partitioned into $256$ non-overlapping intervals $\mathcal{R} = \{R_0, R_1, \dots, R_{255}\}$, with each $R_k \subset \mathbb{R}$.
    \item All feature values in $R_k$ are mapped to the discrete value $k$ via $T$.
    \item For each $k$, we define an inverse transformed value $c_k \in \mathbb{R}$.
\end{itemize}
The forward transformation is defined as:
\begin{equation}
    T(f_i) = k \quad \text{if} \quad f_i \in R_k,
\end{equation}
and the inverse transformation is:
\begin{equation}
    T^{-1}(k) = c_k.
\end{equation}

The objective is to minimize the transformation distortion:
\begin{equation}
    \mathcal{D}_{\text{transform}} = \mathbb{E}_{\mathbf{f} \sim p_f} \left[ \sum_{i=1}^d \left( f_i - T^{-1}({T(f_i)}) \right)^2 \right].
\end{equation}
This optimization is carried out iteratively through two steps:

\paragraph{\textbf{Assignment step:}} Given current inverse transformed values $\{c_k\}$, update each region $R_k$ as:
\begin{equation}
    R_k = \left\{ f_i \in \mathbb{R} \mid k = \arg\min_{j \in \{0,\dots,255\}} \left| f_i - c_j \right| \right\}.
\end{equation}

\paragraph{\textbf{Update step:}} Given updated regions $\{R_k\}$, recompute the inverse transformed values:
\begin{equation}
    c_k = \frac{1}{|R_k|} \sum_{f_i \in R_k} f_i.
\end{equation}

We alternate between these steps until convergence of $\mathcal{D}_{\text{transform}}$. The resulting region-to-integer mapping forms the non-uniform transformation $T(\cdot)$, and $\{c_k\}$ are stored for inverse transform $T^{-1}(\cdot)$. In our experiment, we randomly select 10 features to learn the non-uniform transform.

Figure~\ref{fig_distribution} illustrates the full transformation pipeline of the proposed DT-UFC method. Each row corresponds to a different task, and each column shows a specific stage:

    \paragraph{\textbf{Original Distribution:}} The original feature distribution is often highly skewed or concentrated, as seen in LLaMA3 and DINOv2.
    \paragraph{\textbf{Learned Transform:}} The blue curve plots the learned nonlinear transform function $T(f)$. For peaky distributions, the mapping function becomes steep around highly concentrated regions, assigning more indices to concentrated values. For more uniform inputs (e.g., TTI), the function appears smoother, resembling a near-linear transform.
    \paragraph{\textbf{Region Interval Width:}} We visualize the logarithmic width of each region $R_k$. We observe shorter intervals (denser coverage) near peaks of the original distribution, and longer intervals in sparse regions. The dashed red lines denote the average region interval. 
    \paragraph{\textbf{Transformed Distribution:}} The transformed distribution over integers in $p_t$ becomes much more balanced. The CDF curve is now smooth and monotonic, indicating successful distribution alignment. 
    This transformation not only reduces distributional divergence across models ( Table~\ref{tab_kl}) but also improves latent expressiveness ( See Fig.~\ref{fig_latent}).

In summary, the proposed non-uniform transformation bridges the gap between heterogeneous input distributions and a common encoding space. It enables a universal codec to achieve both high generalization and rate-distortion efficiency across diverse models and tasks.

\begin{table*}
  \caption{Rate-Accuracy Performance on CSR, Seg, and TTI with \textit{Hybrid-Trained Hyperprior} Codecs (Sec. \ref{subsec_rd_performance}, Sec. \ref{subsec_universality})}
  \label{tab_train_hybrid_hyper}
  \resizebox{0.7\textwidth}{!}{%
  \begin{tabular}{@{}c|ccc|ccc|ccc@{}}
\toprule
\textbf{Task}                & \multicolumn{3}{c|}{\textbf{Common Sense Reasoning}} & \multicolumn{3}{c|}{\textbf{Semantic Segmentation}} & \multicolumn{3}{c}{\textbf{Text-to-Image   Synthesis}} \\ \midrule
\textbf{Lambda} & \textbf{BPFP}  & \textbf{Accuracy $\uparrow$}  & \textbf{MSE}  & \textbf{BPFP}   & \textbf{mIoU $\uparrow$}   & \textbf{MSE}   & \textbf{BPFP}     & \textbf{FID $\downarrow$}     & \textbf{MSE}    \\ \midrule
0.0008          & 0.07           & 0.00               & 0.3006        & 0.04            & 34.33           & 3.5556         & 0.05              & 226.74           & 0.1388          \\
0.001           & 0.11           & 0.00               & 0.2836        & 0.07            & 48.83           & 3.3804         & 0.06              & 234.27           & 0.1413          \\
0.0013          & 0.20           & 0.00               & 0.2400        & 0.13            & 60.66           & 4.5479         & 0.07              & 197.31           & 0.1220          \\
0.0018          & 0.49           & 3.80               & 0.1943        & 0.29            & 73.17           & 4.6615         & 0.12              & 140.52           & 0.0902          \\
0.0019          & 0.52           & 4.00               & 0.1892        & 0.31            & 73.85           & 4.2202         & 0.13              & 137.88           & 0.0884          \\
0.0023          & 0.67           & 15.60              & 0.1672        & 0.43            & 75.32           & 2.5743         & 0.16              & 137.83           & 0.0824          \\
0.0025          & 0.95           & 39.80              & 0.1494        & 0.47            & 75.77           & 3.1560         & 0.17              & 127.98           & 0.0788          \\
0.0028          & 1.19           & 72.60              & 0.1342        & 0.54            & 76.95           & 2.8723         & 0.19              & 119.43           & 0.0750          \\
0.003           & 1.23           & 75.40              & 0.1177        & 0.60            & 77.20           & 2.4534         & 0.20              & 117.31           & 0.0707          \\
0.004           & 1.39           & 85.60              & 0.1142        & 0.84            & 78.81           & 1.8782         & 0.27              & 112.95           & 0.0625          \\
0.005           & 1.52           & 97.40              & 0.0904        & 1.00            & 79.83           & 1.8823         & 0.33              & 100.11           & 0.0551          \\
0.007           & 1.68           & 99.40              & 0.0782        & 1.26            & 80.65           & 1.8499         & 0.46              & 80.26            & 0.0441          \\
0.01            & 1.89           & 98.80              & 0.0715        & 1.48            & 80.74           & 1.6486         & 0.71              & 62.06            & 0.0337          \\
0.02            & 2.59           & 98.00              & 0.0454        & 2.03            & 81.72           & 1.0889         & 1.15              & 38.91            & 0.0217          \\ \bottomrule
\end{tabular}}
\end{table*}

\begin{table*}
  \caption{Rate-Accuracy Performance on CSR, Seg, and TTI with \textit{Hybrid-Trained ELIC} Codecs (Sec. \ref{subsec_rd_performance}, Sec. \ref{subsec_universality})}
  \label{tab_train_hybrid_elic}
  \resizebox{0.7\textwidth}{!}{%
  \begin{tabular}{@{}c|ccc|ccc|ccc@{}}
\toprule
\textbf{Task}              & \multicolumn{3}{c|}{\textbf{Common Sense Reasoning}} & \multicolumn{3}{c|}{\textbf{Semantic Segmentation}} & \multicolumn{3}{c}{\textbf{Text-to-Image   Synthesis}} \\ \midrule
\textbf{Lambda} & \textbf{BPFP}  & \textbf{Accuracy $\uparrow$}  & \textbf{MSE}  & \textbf{BPFP}   & \textbf{mIoU $\uparrow$}   & \textbf{MSE}   & \textbf{BPFP}     & \textbf{FID $\downarrow$}     & \textbf{MSE}    \\ \midrule
0.0001          & 0.04           & 0.00               & 0.3316        & 0.005           & 15.23           & 7.0575         & 0.02              & 226.86           & 0.1751          \\
0.0003          & 0.21           & 0.00               & 0.2986        & 0.06            & 70.02           & 3.0153         & 0.04              & 186.87           & 0.1267          \\
0.001           & 0.36           & 3.80               & 0.2199        & 0.16            & 77.17           & 3.0283         & 0.09              & 137.92           & 0.0890          \\
0.0015          & 0.64           & 34.60              & 0.1702        & 0.28            & 78.28           & 2.6021         & 0.12              & 118.64           & 0.0765          \\
0.0019          & 0.70           & 46.80              & 0.1579        & 0.33            & 78.59           & 2.5045         & 0.14              & 112.00           & 0.0716          \\
0.0021          & 0.94           & 69.40              & 0.1433        & 0.47            & 78.92           & 2.3994         & 0.16              & 101.74           & 0.0663          \\
0.0025          & 1.02           & 75.40              & 0.1332        & 0.53            & 79.18           & 2.4558         & 0.18              & 95.59            & 0.0625          \\
0.003           & 1.21           & 93.40              & 0.1125        & 0.61            & 79.24           & 2.2981         & 0.21              & 88.92            & 0.0586          \\
0.004           & 1.65           & 94.00              & 0.0875        & 0.75            & 79.78           & 2.0942         & 0.28              & 76.84            & 0.0507          \\
0.005           & 1.90           & 97.00              & 0.0641        & 0.94            & 80.13           & 1.9586         & 0.37              & 66.24            & 0.0440          \\
0.007           & 2.21           & 94.20              & 0.0577        & 1.08            & 81.35           & 1.3101         & 0.44              & 63.32            & 0.0399          \\
0.01            & 2.51           & 98.60              & 0.0464        & 1.58            & 81.41           & 1.2771         & 0.59              & 54.21            & 0.0327          \\
0.015           & 2.97           & 99.20              & 0.0317        & 1.79            & 81.64           & 1.2243         & 0.87              & 31.37            & 0.0225          \\
0.02            & 3.16           & 98.80              & 0.0192        & 2.00            & 82.73           & 0.6090         & 1.06              & 21.85            & 0.0174          \\ \bottomrule
\end{tabular}}
\end{table*}

\begin{figure*}[h]
  \centering
    \resizebox{0.9\textwidth}{!}{%
    \begin{tabular}{ccc}
        \includegraphics[height=4cm]{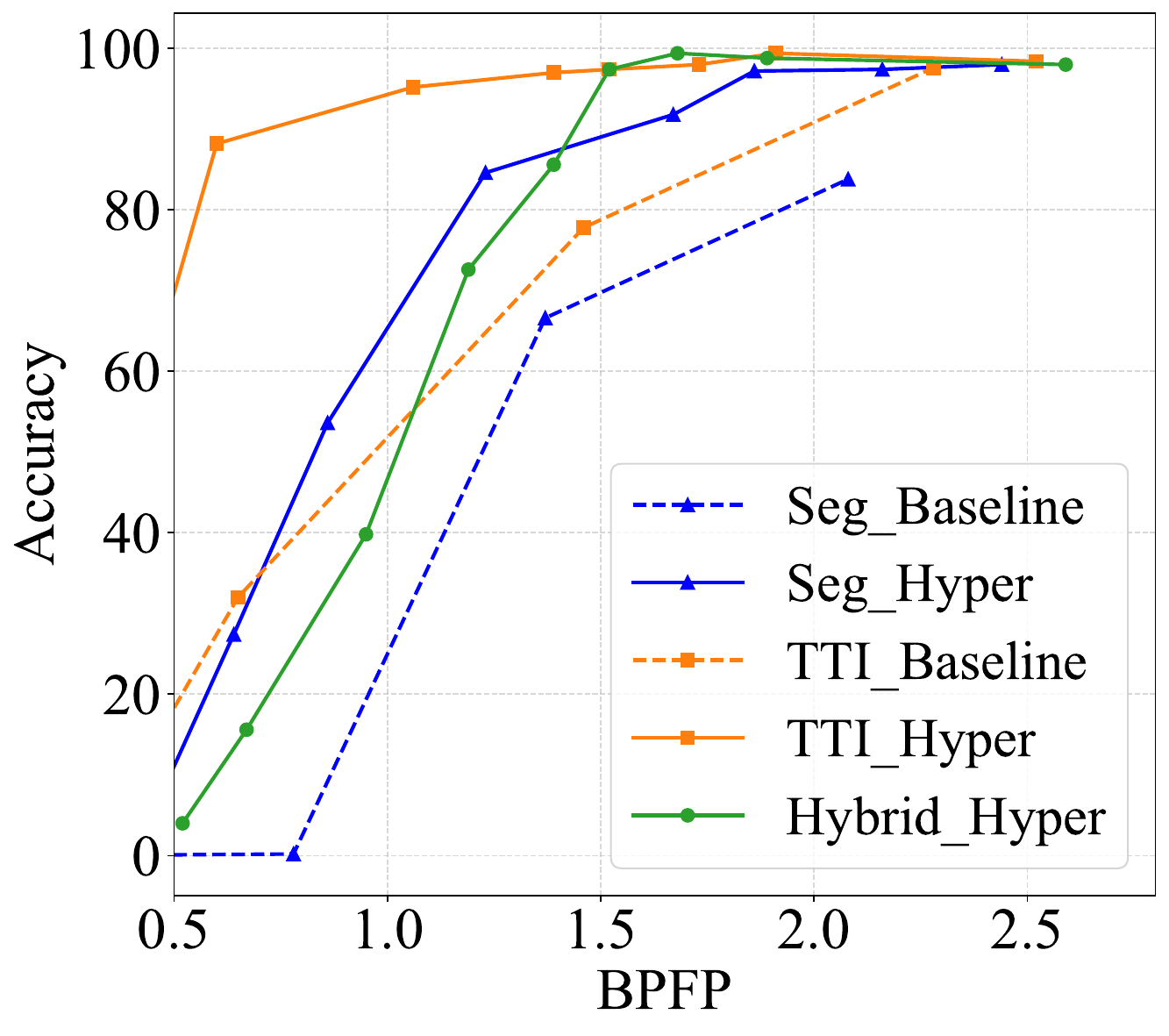} &
        \includegraphics[height=4.08cm]{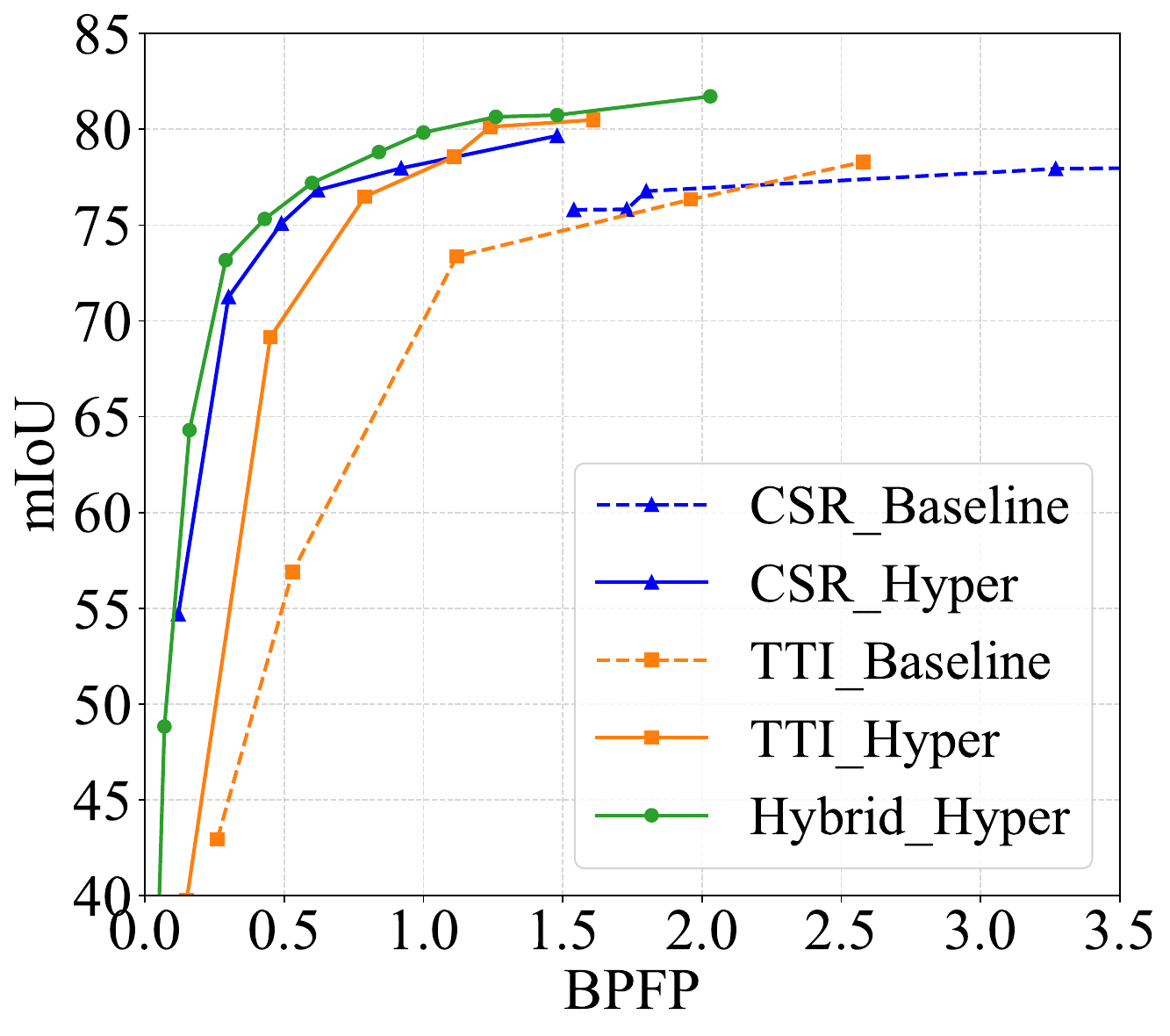} &
        \includegraphics[height=4cm]{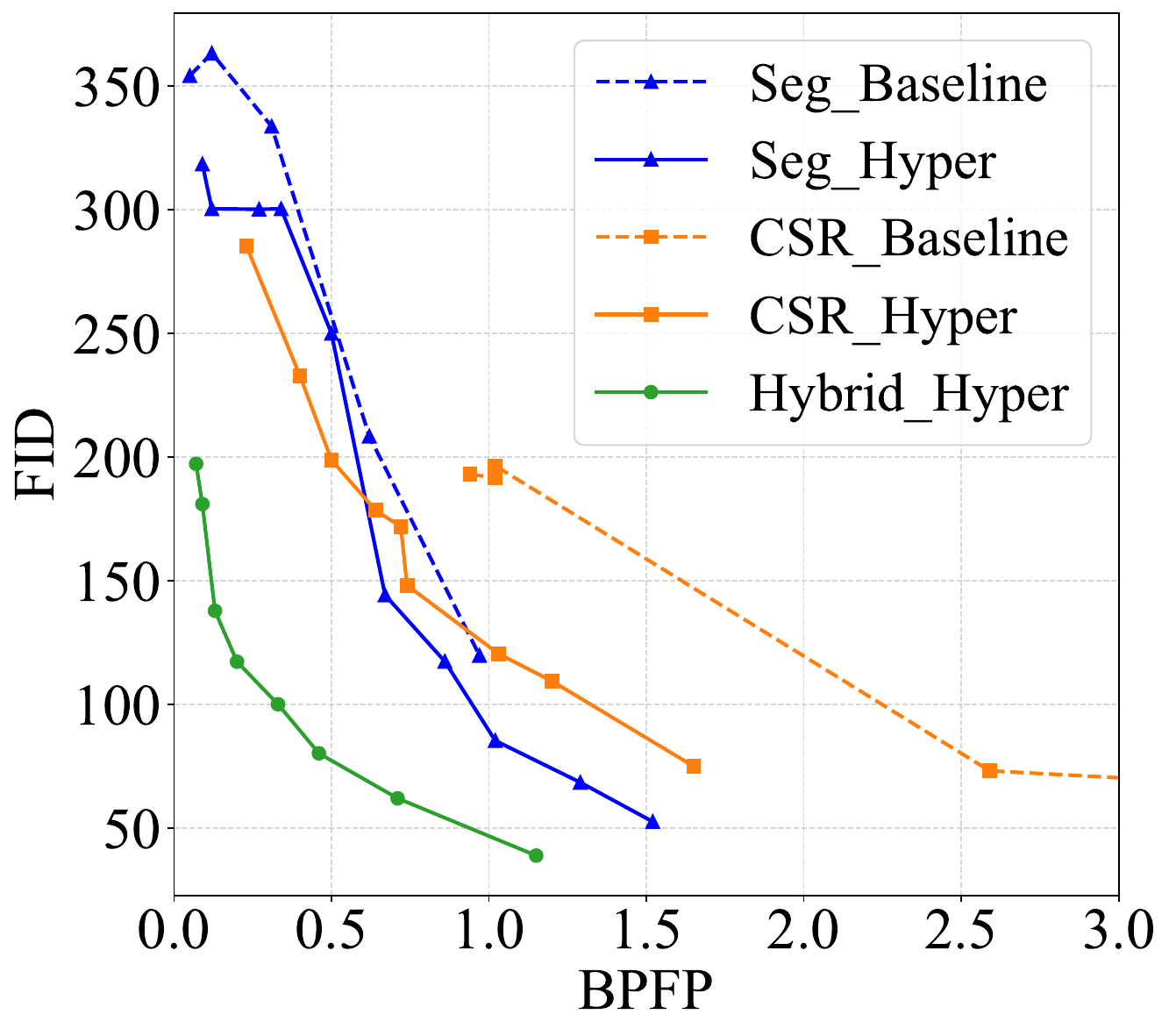} \\
    \end{tabular}
    }
  \caption{Universality comparison between the baseline and the proposed method. The left, middle, and right figures correspond to CSR, Seg, and TTI, respectively. (Sec. \ref{subsec_universality})}
  \label{fig_universality}
\end{figure*}

\begin{table}
  \caption{Rate-Accuracy Performance on CSR, Seg, and TTI with \textit{Task-Specifically Trained Hyperprior} Codecs}
  \label{tab_train_separate}
  \resizebox{\columnwidth}{!}{%
  \begin{tabular}{@{}c|c|cc|cc|cc@{}}
\toprule
\textbf{}                             & \textbf{Task}       & \multicolumn{2}{c|}{\textbf{CSR}} & \multicolumn{2}{c|}{\textbf{Seg}} & \multicolumn{2}{c}{\textbf{TTI}} \\ \midrule
\textbf{Metric}                           & \textbf{Lambda} & \textbf{BPFP}   & \textbf{Acc.}  & \textbf{BPFP}   & \textbf{mIoU}  & \textbf{BPFP}   & \textbf{FID}   \\ \midrule
\multirow{9}{*}{\textbf{\begin{tabular}[c]{@{}c@{}}CSR\\Models\end{tabular}}}
  & 0.0005          & 0.11            & 0.00           & 0.05            & 10.61          & 0.13            & 327.38         \\
                                          & 0.0008          & 0.30            & 0.00           & 0.12            & 54.69          & 0.23            & 285.36         \\
                                          & 0.0017          & 0.68            & 36.00          & 0.39            & 73.10          & 0.50            & 198.86         \\
                                          & 0.0019          & 0.87            & 44.40          & 0.49            & 75.09          & 0.64            & 178.49         \\
                                          & 0.002           & 0.94            & 59.00          & 0.52            & 75.62          & 0.72            & 171.93         \\
                                          & 0.0025          & 1.05            & 79.60          & 0.62            & 76.83          & 0.74            & 148.21         \\
                                          & 0.003           & 1.56            & 97.80          & 0.92            & 77.97          & 1.03            & 120.52         \\
                                          & 0.0035          & 1.74            & 97.40          & 1.03            & 77.84          & 1.20            & 109.46         \\
                                          & 0.006           & 2.23            & 98.00          & 1.48            & 79.66          & 1.65            & 75.02          \\ \midrule
\multirow{12}{*}{\textbf{\begin{tabular}[c]{@{}c@{}}Seg\\Models\end{tabular}}} & 0.0007          & 0.17            & 0.00           & 0.05            & 50.67          & 0.09            & 318.53         \\
                                          & 0.0008          & 0.21            & 0.00           & 0.08            & 60.69          & 0.12            & 300.42         \\
                                          & 0.002           & 0.46            & 6.40           & 0.21            & 71.89          & 0.27            & 300.25         \\
                                          & 0.0025          & 0.64            & 27.40          & 0.29            & 74.76          & 0.34            & 300.40         \\
                                          & 0.003           & 0.86            & 53.60          & 0.42            & 78.10          & 0.50            & 250.03         \\
                                          & 0.004           & 1.23            & 84.60          & 0.65            & 79.09          & 0.67            & 144.28         \\
                                          & 0.005           & 1.67            & 91.80          & 0.94            & 79.59          & 0.86            & 117.41         \\
                                          & 0.006           & 1.86            & 97.20          & 1.10            & 80.05          & 1.01            & 111.02         \\
                                          & 0.01            & 2.16            & 97.40          & 1.41            & 81.02          & 1.29            & 68.50          \\
                                          & 0.015           & 2.44            & 98.00          & 1.65            & 82.02          & 1.52            & 52.64          \\ \midrule
\multirow{9}{*}{\textbf{\begin{tabular}[c]{@{}c@{}}TTI\\Models\end{tabular}}}  & 0.001           & 0.09            & 0.00           & 0.06            & 28.64          & 0.06            & 183.68         \\
                                          & 0.002           & 0.21            & 16.20          & 0.15            & 39.78          & 0.11            & 136.78         \\
                                          & 0.004           & 0.60            & 88.20          & 0.45            & 69.15          & 0.22            & 100.26         \\
                                          & 0.006           & 1.06            & 95.20          & 0.79            & 76.49          & 0.33            & 83.42          \\
                                          & 0.008           & 1.39            & 97.00          & 1.11            & 78.56          & 0.46            & 67.20          \\
                                          & 0.01            & 1.52            & 97.40          & 1.24            & 80.13          & 0.56            & 58.27          \\
                                          & 0.015           & 1.73            & 98.00          & 1.43            & 78.62          & 0.77            & 47.79          \\
                                          & 0.02            & 1.91            & 99.40          & 1.61            & 80.49          & 0.92            & 38.69          \\
                                          & 0.05            & 2.52            & 98.40          & 2.16            & 80.62          & 1.44            & 22.16          \\ \bottomrule
\end{tabular}}
\end{table}

\section{Experiments}
\subsection{Dataset}
For each task, we first extract 10,000 original features from pre-trained large models to construct the training set. These features are then transformed using the proposed non-uniform transformation to generate training inputs for the codec.

For evaluation, we follow the unified test condition proposed in \cite{gao2024feature}, using 100, 500, and 500 features for the Seg, CSR, and TTI tasks, respectively. All features are packed using the same method in \cite{gao2024feature}.

\subsection{Codec and Training Details}
We adopt two representative feature codecs: Hyperprior \cite{balle2018variational} and ELIC \cite{henzel2022efficient}. 
To investigate the impact of training data on performance, we consider two strategies: hybrid training and task-specific training. In the hybrid training setting, all 30000 features across tasks are used. In the task-specific training setting, only the 10000 features corresponding to each task are used for that task.

The feature codecs are optimized using the following loss function:
\begin{equation}
L =  BPFP + \lambda \times ||(F_{o} - \hat{F_{r}})||^2
\end{equation}
where BPFP measures the bitrate. $\lambda$ is a scaling factor used to adjust the bitrate. $F_o$ and $F_r$ denote the original input feature and reconstructed feature, respectively.
Both training strategies use the same training configuration. We employ the Adam optimizer with an initial learning rate of 1e-4, which is adjusted by the ReduceLROnPlateau scheduler. Training continues until the learning rate drops to 1e-8, followed by 20 additional epochs. The model with the lowest validation loss is selected. 
To control the bitrate, we adjust the Lagrange multiplier $\lambda$. Detailed settings are provided in Tables \ref{tab_train_hybrid_hyper}, \ref{tab_train_hybrid_elic}, and \ref{tab_train_separate}. All experiments are conducted on Nvidia RTX 3090 GPUs.

\subsection{Rate-Accuracy Performance}
\label{subsec_rd_performance}
This subsection validates the performance of DT-UFC in terms of rate-accuracy across three representative tasks: CSR, Seg, and TTI. We compare DT-UFC against baseline methods in both the task-specifically trained and hybrid-trained strategies. Fig. \ref{fig_rd} shows the rate-accuracy curves, with numerical results in Tables \ref{tab_train_hybrid_hyper}, \ref{tab_train_hybrid_elic}, and \ref{tab_train_separate}. For clarity, some points are omitted from the figures.

In both CSR and Seg tasks, DT-UFC consistently outperforms the baselines under both training strategies, especially at higher bitrates. This demonstrates that our proposed feature transformation effectively addresses the issue of peaky and concentrated feature distributions. Generally, the CSR task exhibits a higher correlation between the feature MSE and task accuracy than the Seg task. 
We attribute this to the broader original feature distribution range observed in the Seg task, which introduces greater nonlinearity in the rate–accuracy relationship.

In the TTI task, while the hybrid-trained ELIC codecs outperform both baselines consistently, the hybrid-trained Hyperprior codecs slightly underperform their task-specifically trained counterparts at low bitrates. This indicates that TTI features, which are most similar to image distributions, benefit more from high-capacity codecs like ELIC. 

Additionally, in the CSR task, the task-specifically trained Hyperprior codec achieves better performance than the more sophisticated ELIC codec. This phenomenon reveals an inherent distinction between textual and visual feature distributions: codecs that excel on visual features may not perform well on textual features and may even hurt performance.

In summary, the proposed distribution transformation provides a more encoder-friendly feature representation space that enables both lightweight and high-capacity codecs to operate effectively. This is the key to realizing high-performance, task-agnostic feature coding.

\subsection{Cross-Task Universality}
\label{subsec_universality}
To evaluate the universality of DT-UFC, we test its ability to compress features from unseen tasks using models trained on a different task. Fig. \ref{fig_universality} shows the rate-accuracy performance across tasks.

In the CSR task, baseline models trained on Seg or TTI perform poorly as they fail to generalize to the highly distinct textual CSR features. In contrast, DT-UFC achieves strong performance even when trained on other tasks, demonstrating the effectiveness of our transformation in aligning heterogeneous distributions. By transforming task-specific features into a unified space, DT-UFC reduces distributional divergence and enables successful cross-task compression.

In the Seg task, the baseline CSR models fail to achieve low bitrates, while all the DT-UFC models achieve smooth rate-accuracy curves. For the TTI models, both hybrid-trained and task-specifically trained DT-UFC models outperform their baseline counterparts by a large margin, indicating higher robustness of DT-UFC in the TTI task.

In the TTI task, all the baseline models and task-specifically trained models show unstable rate-accuracy behaviors.
Nevertheless, DT-UFC still outperforms the baselines, especially when trained on CSR. The hybrid-trained variant demonstrates notable robustness and achieves the best overall performance.

Furthermore, comparing task-specific and hybrid training reveals that hybrid training generally improves generalization, especially in Seg and TTI tasks, by exposing the codec to more diverse feature distributions. However, in the CSR task, hybrid training slightly degrades performance at low bitrates, possibly due to the model's adaptation to visual features. This highlights the importance of balancing feature diversity in multi-task scenarios.

Overall, DT-UFC's distribution alignment strategy ensures reliable encoder performance across tasks, even in zero-shot scenarios. This confirms its potential as a universal feature coding solution across tasks, models, and modalities.

\begin{table}
  \caption{Rate-Accuracy Performance Comparison between the Baseline, Baseline with Transformed Features, and DT-UFC with Baseline Features (Sec. \ref{subsec_transform} and Sec. \ref{subsec_codec})}
  \label{tab_transform_codec}
  \resizebox{0.9\columnwidth}{!}{%
  \begin{tabular}{@{}c|cc|cc|cc@{}}
    \toprule
    \textbf{Task} & \multicolumn{2}{c|}{\textbf{CSR}} & \multicolumn{2}{c|}{\textbf{Seg}} & \multicolumn{2}{c}{\textbf{TTI}} \\
    \midrule
    \textbf{Metric} & \textbf{BPFP} & \textbf{Acc.} & \textbf{BPFP} & \textbf{mIoU} & \textbf{BPFP} & \textbf{FID} \\
    \midrule
    \textbf{Original} & \textbf{32} & \textbf{100} & \textbf{32} & \textbf{83.39} & \textbf{32} & \textbf{0} \\
    \midrule
    \multirow{5}{*}{\textbf{Baseline}} 
    & 1.35 & 34.80 & 0.03 & 37.11 & 0.08 & 170.01 \\
    & 1.50 & 57.80 & 0.12 & 62.58 & 0.15 & 124.84 \\
    & 1.68 & 82.40 & 0.54 & 74.82 & 0.28 & 95.03 \\
    & 3.60 & 87.80 & 1.30 & 77.27 & 0.66 & 52.12 \\
    & 6.34 & 91.40 & 1.71 & 77.96 & 1.44 & 25.25 \\
    \midrule
    \multirow{5}{*}{\textbf{\begin{tabular}[c]{@{}c@{}c@{}c@{}}Baseline with \\ Transformed \\ Features \\ (Sec. \ref{subsec_transform}) \end{tabular}}} 
    & 2.08 & 94.00 & 0.04 & 31.60 & 0.12 & 161.72 \\
    & 2.34 & 91.40 & 0.12 & 64.60 & 0.26 & 115.86 \\
    & 2.49 & 92.00 & 0.51 & 77.72 & 0.56 & 80.91 \\
    & 3.91 & 96.00 & 1.23 & 80.14 & 1.18 & 35.26 \\
    & 7.02 & 96.80 & 1.65 & 81.34 & 1.93 & 22.60 \\
    \midrule
    \multirow{5}{*}{\textbf{\begin{tabular}[c]{@{}c@{}c@{}c@{}}DT-UFC with \\ Baseline \\ Features \\ (Sec. \ref{subsec_codec}) \end{tabular}}}
    & 0.30 & 6.60  & 0.05 & 37.32 & 0.08 & 166.57 \\
    & 0.35 & 14.80 & 0.17 & 62.41 & 0.14 & 144.11 \\
    & 0.50 & 20.80 & 0.45 & 73.50 & 0.18 & 128.57 \\
    & 0.97 & 71.00 & 1.09 & 76.55 & 0.33 & 96.84  \\
    & 1.41 & 97.20 & 1.55 & 77.94 & 0.54 & 72.97  \\
    \bottomrule
  \end{tabular}
}
\end{table}
\begin{table}
  \caption{Rate-Accuracy Performance of DT-UFC on CNN Features (Sec. \ref{subsec_architecture})}
  \label{tab_cnn_cls}
  \resizebox{0.55\columnwidth}{!}{%
  \begin{tabular}{@{}cccc@{}}
    \toprule
    \textbf{Lambda} & \textbf{BPFP} & \textbf{Accuracy} & \textbf{MSE} \\ \midrule
    \textbf{Original}         & \textbf{32}         & \textbf{86.40 }            & \textbf{0}       \\ \midrule
    0.0005          & 0.06          & 84.40             & 0.5006       \\
    0.0008          & 0.09          & 85.00             & 0.4627       \\
    0.0015          & 0.19          & 85.80             & 0.3350       \\
    0.0025          & 0.43          & 86.00             & 0.1749       \\
    0.004           & 0.88          & 86.40             & 0.0950       \\
    \bottomrule
  \end{tabular}}
\end{table}
\subsection{Universality of Transformed Features}
\label{subsec_transform}
To isolate the contribution of the distribution transformation, we directly feed transformed features into the baseline codecs without any retraining and compare their rate-accuracy performance with the baseline. 

As shown in Table \ref{tab_transform_codec}, this simple substitution leads to significant improvements across all tasks. In the CSR task, the baseline codecs consistently encode the transformed feature with high bitrates yet retain high accuracy. In the Seg and TTI tasks, the trends in bitrate and accuracy remain similar to the baseline, indicating no degradation.
These results demonstrate that the transformation alone enhances feature compressibility by reshaping feature distributions. Even without retraining the codec, the benefits are clearly observed.
In essence, the transformation serves not only as an enabler for universal encoder training but also as a standalone mechanism for improving structural expressiveness and compressibility of features.

\subsection{Codec's Robustness to Imperfect Inputs}
\label{subsec_codec}
We further examine the robustness of DT-UFC codecs when given imperfect input. Specifically, we feed features truncated by baseline pre-processing to the hybrid-trained feature codecs. 

As reported in Table \ref{tab_transform_codec}, DT-UFC codecs maintain a competitive performance under these degraded conditions.
For the CSR task, DT-UFC substantially outperforms the baseline, achieving over 40\% accuracy improvement at 1.5 BPFP. In Seg and TTI tasks, DT-UFC maintains comparable performance with the baseline. This demonstrates that even when inputs deviate from the ideal transformed space, codecs trained using the transformed features still maintains their effectiveness.

Moreover, the decline of metrics such as MSE and FID across bitrate indicates that codecs trained with transformed features have not merely memorized specific input patterns but have learned to model underlying feature structures more generally.
Therefore, DT-UFC is not only effective under ideal conditions but also robust in real-world deployment scenarios where features may be distorted or misaligned. This resilience is critical for practical adoption.

\subsection{Architecture-Agnostic Generalization}
\label{subsec_architecture}
To assess cross-architecture generalization, we apply DT-UFC to CNN-derived features. Specifically, we extract features from ResNet-50 on an image classification task. A transformation is learned on 10 features and applied before encoding. Instead of training feature codecs on CNN features, we feed them to the hybrid-trained Hyperprior codecs.

Table \ref{tab_cnn_cls} shows that DT-UFC achieves nearly lossless classification accuracy and decreasing MSE across bitrates. This demonstrates the codec's strong generalization from Transformer features to CNN features.
Importantly, this result highlights our method’s ability to align feature distributions across architectural paradigms, enabling universal feature compression across diverse model backbones.
In addition, the generalization ability also indicates that the proposed transformation can align feature distributions across model architectures, enabling universal feature coding across Transformer-based and CNN-based backbones. The strong performance without retraining further validates the scalability and practicality of DT-UFC as a truly architecture-agnostic feature codec.

\section{Conclusion}
In this paper, we present DT-UFC, a novel framework for universal feature coding across large models by introducing a peaky-to-balanced distribution transformation. 
By aligning diverse feature distributions into a common, balanced space, DT-UFC enables a single codec to generalize across models and tasks without retraining.
Extensive experiments demonstrate that DT-UFC not only improves rate-distortion performance over task-specific baselines, but also supports architecture-agnostic and cross-task generalization, confirming its practical value in large model deployment scenarios.

In future work, we aim to integrate DT-UFC into real-world distributed AI systems to evaluate its impact on latency, energy efficiency, and scalability.
We hope this work inspires further research on universal and practical feature coding.

\begin{acks}
This work was supported by the Ministry of Education of Singapore under Grant T2EP20123-0006. We acknowledge the support of GPU cluster built by MCC Lab of Information Science and Technology Institution, USTC.
\end{acks}

\newpage
\bibliographystyle{ACM-Reference-Format}
\bibliography{refs}


\begin{thebibliography}{69}


\ifx \showCODEN    \undefined \def \showCODEN     #1{\unskip}     \fi
\ifx \showISBNx    \undefined \def \showISBNx     #1{\unskip}     \fi
\ifx \showISBNxiii \undefined \def \showISBNxiii  #1{\unskip}     \fi
\ifx \showISSN     \undefined \def \showISSN      #1{\unskip}     \fi
\ifx \showLCCN     \undefined \def \showLCCN      #1{\unskip}     \fi
\ifx \shownote     \undefined \def \shownote      #1{#1}          \fi
\ifx \showarticletitle \undefined \def \showarticletitle #1{#1}   \fi
\ifx \showURL      \undefined \def \showURL       {\relax}        \fi
\providecommand\bibfield[2]{#2}
\providecommand\bibinfo[2]{#2}
\providecommand\natexlab[1]{#1}
\providecommand\showeprint[2][]{arXiv:#2}

\bibitem[Achiam et~al\mbox{.}(2023)]%
        {achiam2023gpt}
\bibfield{author}{\bibinfo{person}{Josh Achiam}, \bibinfo{person}{Steven Adler}, \bibinfo{person}{Sandhini Agarwal}, \bibinfo{person}{Lama Ahmad}, \bibinfo{person}{Ilge Akkaya}, \bibinfo{person}{Florencia~Leoni Aleman}, \bibinfo{person}{Diogo Almeida}, \bibinfo{person}{Janko Altenschmidt}, \bibinfo{person}{Sam Altman}, \bibinfo{person}{Shyamal Anadkat}, {et~al\mbox{.}}} \bibinfo{year}{2023}\natexlab{}.
\newblock \showarticletitle{Gpt-4 technical report}.
\newblock \bibinfo{journal}{\emph{arXiv preprint arXiv:2303.08774}} (\bibinfo{year}{2023}).
\newblock


\bibitem[Alvar and Bajić(2019)]%
        {alvar2019multi}
\bibfield{author}{\bibinfo{person}{Saeed~Ranjbar Alvar} {and} \bibinfo{person}{Ivan~V. Bajić}.} \bibinfo{year}{2019}\natexlab{}.
\newblock \showarticletitle{Multi-Task Learning with Compressible Features for Collaborative Intelligence}. In \bibinfo{booktitle}{\emph{ICIP}}. \bibinfo{pages}{1705--1709}.
\newblock
\href{https://doi.org/10.1109/ICIP.2019.8803110}{doi:\nolinkurl{10.1109/ICIP.2019.8803110}}


\bibitem[Alvar and Bajić(2020)]%
        {alvar2020bit}
\bibfield{author}{\bibinfo{person}{Saeed~Ranjbar Alvar} {and} \bibinfo{person}{Ivan~V. Bajić}.} \bibinfo{year}{2020}\natexlab{}.
\newblock \showarticletitle{Bit Allocation for Multi-Task Collaborative Intelligence}. In \bibinfo{booktitle}{\emph{ICASSP}}. \bibinfo{pages}{4342--4346}.
\newblock
\href{https://doi.org/10.1109/ICASSP40776.2020.9054770}{doi:\nolinkurl{10.1109/ICASSP40776.2020.9054770}}


\bibitem[Arpit et~al\mbox{.}(2017)]%
        {arpit2017closer}
\bibfield{author}{\bibinfo{person}{Devansh Arpit}, \bibinfo{person}{Stanis\l{}aw Jastrzundefinedbski}, \bibinfo{person}{Nicolas Ballas}, \bibinfo{person}{David Krueger}, \bibinfo{person}{Emmanuel Bengio}, \bibinfo{person}{Maxinder~S. Kanwal}, \bibinfo{person}{Tegan Maharaj}, \bibinfo{person}{Asja Fischer}, \bibinfo{person}{Aaron Courville}, \bibinfo{person}{Yoshua Bengio}, {and} \bibinfo{person}{Simon Lacoste-Julien}.} \bibinfo{year}{2017}\natexlab{}.
\newblock \showarticletitle{A closer look at memorization in deep networks} \emph{(\bibinfo{series}{ICML'17})}. \bibinfo{publisher}{JMLR}, \bibinfo{pages}{233–242}.
\newblock


\bibitem[Bai et~al\mbox{.}(2023)]%
        {bai2023qwen}
\bibfield{author}{\bibinfo{person}{Jinze Bai}, \bibinfo{person}{Shuai Bai}, \bibinfo{person}{Yunfei Chu}, \bibinfo{person}{Zeyu Cui}, \bibinfo{person}{Kai Dang}, \bibinfo{person}{Xiaodong Deng}, \bibinfo{person}{Yang Fan}, \bibinfo{person}{Wenbin Ge}, \bibinfo{person}{Yu Han}, \bibinfo{person}{Fei Huang}, {et~al\mbox{.}}} \bibinfo{year}{2023}\natexlab{}.
\newblock \showarticletitle{Qwen technical report}.
\newblock \bibinfo{journal}{\emph{arXiv preprint arXiv:2309.16609}} (\bibinfo{year}{2023}).
\newblock


\bibitem[Ball{\'e} et~al\mbox{.}(2018)]%
        {balle2018variational}
\bibfield{author}{\bibinfo{person}{Johannes Ball{\'e}}, \bibinfo{person}{David~C. Minnen}, \bibinfo{person}{Saurabh Singh}, \bibinfo{person}{Sung~Jin Hwang}, {and} \bibinfo{person}{Nick Johnston}.} \bibinfo{year}{2018}\natexlab{}.
\newblock \showarticletitle{Variational image compression with a scale hyperprior}.
\newblock \bibinfo{journal}{\emph{ArXiv}}  \bibinfo{volume}{abs/1802.01436} (\bibinfo{year}{2018}).
\newblock


\bibitem[Cai et~al\mbox{.}(2022)]%
        {cai2022high}
\bibfield{author}{\bibinfo{person}{Yangang Cai}, \bibinfo{person}{Peiyin Xing}, {and} \bibinfo{person}{Xuesong Gao}.} \bibinfo{year}{2022}\natexlab{}.
\newblock \showarticletitle{High Efficient {3D} Convolution Feature Compression}.
\newblock \bibinfo{journal}{\emph{IEEE Transactions on Circuits and Systems for Video Technology}} (\bibinfo{year}{2022}), \bibinfo{pages}{1--1}.
\newblock
\href{https://doi.org/10.1109/TCSVT.2022.3200698}{doi:\nolinkurl{10.1109/TCSVT.2022.3200698}}


\bibitem[Chen et~al\mbox{.}(2024)]%
        {chen2024end}
\bibfield{author}{\bibinfo{person}{Qiaoxi Chen}, \bibinfo{person}{Changsheng Gao}, {and} \bibinfo{person}{Dong Liu}.} \bibinfo{year}{2024}\natexlab{}.
\newblock \showarticletitle{End-to-End Learned Scalable Multilayer Feature Compression For Machine Vision Tasks}. In \bibinfo{booktitle}{\emph{ICIP}}. \bibinfo{pages}{1781--1787}.
\newblock
\href{https://doi.org/10.1109/ICIP51287.2024.10647798}{doi:\nolinkurl{10.1109/ICIP51287.2024.10647798}}


\bibitem[Chen et~al\mbox{.}(2023)]%
        {chen2023transtic}
\bibfield{author}{\bibinfo{person}{Yi-Hsin Chen}, \bibinfo{person}{Ying-Chieh Weng}, \bibinfo{person}{Chia-Hao Kao}, \bibinfo{person}{Cheng Chien}, \bibinfo{person}{Wei-Chen Chiu}, {and} \bibinfo{person}{Wen-Hsiao Peng}.} \bibinfo{year}{2023}\natexlab{}.
\newblock \showarticletitle{{TransTIC}: Transferring transformer-based image compression from human perception to machine perception}. In \bibinfo{booktitle}{\emph{Proceedings of the IEEE/CVF International Conference on Computer Vision}}. \bibinfo{pages}{23297--23307}.
\newblock


\bibitem[Chen et~al\mbox{.}(2019)]%
        {chen2019lossy}
\bibfield{author}{\bibinfo{person}{Zhuo Chen}, \bibinfo{person}{Kui Fan}, \bibinfo{person}{Shiqi Wang}, \bibinfo{person}{Ling-Yu Duan}, \bibinfo{person}{Weisi Lin}, {and} \bibinfo{person}{Alex Kot}.} \bibinfo{year}{2019}\natexlab{}.
\newblock \showarticletitle{Lossy Intermediate Deep Learning Feature Compression and Evaluation}. In \bibinfo{booktitle}{\emph{Proceedings of the 27th ACM International Conference on Multimedia}} \emph{(\bibinfo{series}{MM '19})}. \bibinfo{publisher}{Association for Computing Machinery}, \bibinfo{address}{New York, NY, USA}, \bibinfo{pages}{2414–2422}.
\newblock
\showISBNx{9781450368896}
\href{https://doi.org/10.1145/3343031.3350849}{doi:\nolinkurl{10.1145/3343031.3350849}}


\bibitem[Choi and Bajić(2018)]%
        {choi2018deep}
\bibfield{author}{\bibinfo{person}{Hyomin Choi} {and} \bibinfo{person}{Ivan~V. Bajić}.} \bibinfo{year}{2018}\natexlab{}.
\newblock \showarticletitle{Deep Feature Compression for Collaborative Object Detection}. In \bibinfo{booktitle}{\emph{ICIP}}. \bibinfo{pages}{3743--3747}.
\newblock
\href{https://doi.org/10.1109/ICIP.2018.8451100}{doi:\nolinkurl{10.1109/ICIP.2018.8451100}}


\bibitem[Esser et~al\mbox{.}(2024)]%
        {esser2024sd3}
\bibfield{author}{\bibinfo{person}{Patrick Esser}, \bibinfo{person}{Sumith Kulal}, \bibinfo{person}{Andreas Blattmann}, \bibinfo{person}{Rahim Entezari}, \bibinfo{person}{Jonas M{\"u}ller}, \bibinfo{person}{Harry Saini}, \bibinfo{person}{Yam Levi}, \bibinfo{person}{Dominik Lorenz}, \bibinfo{person}{Axel Sauer}, \bibinfo{person}{Frederic Boesel}, {et~al\mbox{.}}} \bibinfo{year}{2024}\natexlab{}.
\newblock \showarticletitle{Scaling rectified flow transformers for high-resolution image synthesis}. In \bibinfo{booktitle}{\emph{ICML}}.
\newblock


\bibitem[Feng et~al\mbox{.}(2022)]%
        {feng2022image}
\bibfield{author}{\bibinfo{person}{Ruoyu Feng}, \bibinfo{person}{Xin Jin}, \bibinfo{person}{Zongyu Guo}, \bibinfo{person}{Runsen Feng}, \bibinfo{person}{Yixin Gao}, \bibinfo{person}{Tianyu He}, \bibinfo{person}{Zhizheng Zhang}, \bibinfo{person}{Simeng Sun}, {and} \bibinfo{person}{Zhibo Chen}.} \bibinfo{year}{2022}\natexlab{}.
\newblock \showarticletitle{Image coding for machines with omnipotent feature learning}. In \bibinfo{booktitle}{\emph{ECCV}}. Springer, \bibinfo{pages}{510--528}.
\newblock


\bibitem[Gao et~al\mbox{.}(2024a)]%
        {gao2024dmofc}
\bibfield{author}{\bibinfo{person}{Changsheng Gao}, \bibinfo{person}{Yiheng Jiang}, \bibinfo{person}{Li Li}, \bibinfo{person}{Dong Liu}, {and} \bibinfo{person}{Feng Wu}.} \bibinfo{year}{2024}\natexlab{a}.
\newblock \showarticletitle{{DMOFC:} Discrimination Metric-Optimized Feature Compression}. In \bibinfo{booktitle}{\emph{PCS}}. \bibinfo{pages}{1--5}.
\newblock
\href{https://doi.org/10.1109/PCS60826.2024.10566361}{doi:\nolinkurl{10.1109/PCS60826.2024.10566361}}


\bibitem[Gao et~al\mbox{.}(2025)]%
        {gao2024imofc}
\bibfield{author}{\bibinfo{person}{Changsheng Gao}, \bibinfo{person}{Yiheng Jiang}, \bibinfo{person}{Siqi Wu}, \bibinfo{person}{Yifan Ma}, \bibinfo{person}{Li Li}, {and} \bibinfo{person}{Dong Liu}.} \bibinfo{year}{2025}\natexlab{}.
\newblock \showarticletitle{{IMOFC:} Identity-Level Metric Optimized Feature Compression for Identification Tasks}.
\newblock \bibinfo{journal}{\emph{IEEE Transactions on Circuits and Systems for Video Technology}} \bibinfo{volume}{35}, \bibinfo{number}{2} (\bibinfo{year}{2025}), \bibinfo{pages}{1855--1869}.
\newblock
\href{https://doi.org/10.1109/TCSVT.2024.3467124}{doi:\nolinkurl{10.1109/TCSVT.2024.3467124}}


\bibitem[Gao et~al\mbox{.}(2022)]%
        {gao2022twostep}
\bibfield{author}{\bibinfo{person}{Changsheng Gao}, \bibinfo{person}{Li Li}, \bibinfo{person}{Dong Liu}, \bibinfo{person}{Zhibo Chen}, \bibinfo{person}{Weiping Li}, {and} \bibinfo{person}{Feng Wu}.} \bibinfo{year}{2022}\natexlab{}.
\newblock \showarticletitle{Two-Step Fast Mode Decision for Intra Coding of Screen Content}.
\newblock \bibinfo{journal}{\emph{IEEE Transactions on Circuits and Systems for Video Technology}} \bibinfo{volume}{32}, \bibinfo{number}{8} (\bibinfo{year}{2022}), \bibinfo{pages}{5608--5622}.
\newblock


\bibitem[Gao et~al\mbox{.}(2024b)]%
        {gao2024rethinking}
\bibfield{author}{\bibinfo{person}{Changsheng Gao}, \bibinfo{person}{Zhuoyuan Li}, \bibinfo{person}{Li Li}, \bibinfo{person}{Dong Liu}, {and} \bibinfo{person}{Feng Wu}.} \bibinfo{year}{2024}\natexlab{b}.
\newblock \showarticletitle{Rethinking the Joint Optimization in Video Coding for Machines: A Case Study}. In \bibinfo{booktitle}{\emph{DCC}}. \bibinfo{pages}{556--556}.
\newblock


\bibitem[Gao et~al\mbox{.}(2023)]%
        {gao2023towards}
\bibfield{author}{\bibinfo{person}{Changsheng Gao}, \bibinfo{person}{Dong Liu}, \bibinfo{person}{Li Li}, {and} \bibinfo{person}{Feng Wu}.} \bibinfo{year}{2023}\natexlab{}.
\newblock \showarticletitle{Towards Task-Generic Image Compression: A Study of Semantics-Oriented Metrics}.
\newblock \bibinfo{journal}{\emph{IEEE Transactions on Multimedia}}  \bibinfo{volume}{25} (\bibinfo{year}{2023}), \bibinfo{pages}{721--735}.
\newblock
\href{https://doi.org/10.1109/TMM.2021.3130754}{doi:\nolinkurl{10.1109/TMM.2021.3130754}}


\bibitem[Gao et~al\mbox{.}(2024c)]%
        {gao2024feature}
\bibfield{author}{\bibinfo{person}{Changsheng Gao}, \bibinfo{person}{Yifan Ma}, \bibinfo{person}{Qiaoxi Chen}, \bibinfo{person}{Yenan Xu}, \bibinfo{person}{Dong Liu}, {and} \bibinfo{person}{Weisi Lin}.} \bibinfo{year}{2024}\natexlab{c}.
\newblock \showarticletitle{Feature Coding in the Era of Large Models: Dataset, Test Conditions, and Benchmark}.
\newblock \bibinfo{journal}{\emph{arXiv preprint arXiv:2412.04307}} (\bibinfo{year}{2024}).
\newblock


\bibitem[Ge et~al\mbox{.}(2021)]%
        {ge2021structured}
\bibfield{author}{\bibinfo{person}{Xuri Ge}, \bibinfo{person}{Fuhai Chen}, \bibinfo{person}{Joemon~M Jose}, \bibinfo{person}{Zhilong Ji}, \bibinfo{person}{Zhongqin Wu}, {and} \bibinfo{person}{Xiao Liu}.} \bibinfo{year}{2021}\natexlab{}.
\newblock \showarticletitle{Structured multi-modal feature embedding and alignment for image-sentence retrieval}. In \bibinfo{booktitle}{\emph{ACM MM}}. \bibinfo{pages}{5185--5193}.
\newblock


\bibitem[Guo et~al\mbox{.}(2025)]%
        {guo2025deepseek}
\bibfield{author}{\bibinfo{person}{Daya Guo}, \bibinfo{person}{Dejian Yang}, \bibinfo{person}{Haowei Zhang}, \bibinfo{person}{Junxiao Song}, \bibinfo{person}{Ruoyu Zhang}, \bibinfo{person}{Runxin Xu}, \bibinfo{person}{Qihao Zhu}, \bibinfo{person}{Shirong Ma}, \bibinfo{person}{Peiyi Wang}, \bibinfo{person}{Xiao Bi}, {et~al\mbox{.}}} \bibinfo{year}{2025}\natexlab{}.
\newblock \showarticletitle{Deepseek-r1: Incentivizing reasoning capability in llms via reinforcement learning}.
\newblock \bibinfo{journal}{\emph{arXiv preprint arXiv:2501.12948}} (\bibinfo{year}{2025}).
\newblock


\bibitem[Guo et~al\mbox{.}(2023)]%
        {guo2023toward}
\bibfield{author}{\bibinfo{person}{Sha Guo}, \bibinfo{person}{Zhuo Chen}, \bibinfo{person}{Yang Zhao}, \bibinfo{person}{Ning Zhang}, \bibinfo{person}{Xiaotong Li}, {and} \bibinfo{person}{Lingyu Duan}.} \bibinfo{year}{2023}\natexlab{}.
\newblock \showarticletitle{Toward Scalable Image Feature Compression: A Content-Adaptive and Diffusion-Based Approach}. In \bibinfo{booktitle}{\emph{Proceedings of the 31st ACM International Conference on Multimedia}} (Ottawa ON, Canada) \emph{(\bibinfo{series}{MM '23})}. \bibinfo{publisher}{Association for Computing Machinery}, \bibinfo{address}{New York, NY, USA}, \bibinfo{pages}{1431–1442}.
\newblock
\showISBNx{9798400701085}
\href{https://doi.org/10.1145/3581783.3611851}{doi:\nolinkurl{10.1145/3581783.3611851}}


\bibitem[Henzel et~al\mbox{.}(2022)]%
        {henzel2022efficient}
\bibfield{author}{\bibinfo{person}{Robert Henzel}, \bibinfo{person}{Kiran Misra}, {and} \bibinfo{person}{Tianying Ji}.} \bibinfo{year}{2022}\natexlab{}.
\newblock \showarticletitle{Efficient Feature Compression for the Object Tracking Task}. In \bibinfo{booktitle}{\emph{ICIP}}. \bibinfo{pages}{3505--3509}.
\newblock
\href{https://doi.org/10.1109/ICIP46576.2022.9897802}{doi:\nolinkurl{10.1109/ICIP46576.2022.9897802}}


\bibitem[Kao et~al\mbox{.}(2024)]%
        {kao2024bridging}
\bibfield{author}{\bibinfo{person}{Chia-Hao Kao}, \bibinfo{person}{Cheng Chien}, \bibinfo{person}{Yu-Jen Tseng}, \bibinfo{person}{Yi-Hsin Chen}, \bibinfo{person}{Alessandro Gnutti}, \bibinfo{person}{Shao-Yuan Lo}, \bibinfo{person}{Wen-Hsiao Peng}, {and} \bibinfo{person}{Riccardo Leonardi}.} \bibinfo{year}{2024}\natexlab{}.
\newblock \showarticletitle{Bridging compressed image latents and multimodal large language models}.
\newblock \bibinfo{journal}{\emph{arXiv preprint arXiv:2407.19651}} (\bibinfo{year}{2024}).
\newblock


\bibitem[Kim et~al\mbox{.}(2023)]%
        {kim2023end}
\bibfield{author}{\bibinfo{person}{Yeongwoong Kim}, \bibinfo{person}{Hyewon Jeong}, \bibinfo{person}{Janghyun Yu}, \bibinfo{person}{Younhee Kim}, \bibinfo{person}{Jooyoung Lee}, \bibinfo{person}{Se~Yoon Jeong}, {and} \bibinfo{person}{Hui~Yong Kim}.} \bibinfo{year}{2023}\natexlab{}.
\newblock \showarticletitle{End-to-End Learnable Multi-Scale Feature Compression for {VCM}}.
\newblock \bibinfo{journal}{\emph{IEEE Transactions on Circuits and Systems for Video Technology}} (\bibinfo{year}{2023}), \bibinfo{pages}{1--1}.
\newblock
\href{https://doi.org/10.1109/TCSVT.2023.3302858}{doi:\nolinkurl{10.1109/TCSVT.2023.3302858}}


\bibitem[Lee and Lee(2023)]%
        {joonho2023feature}
\bibfield{author}{\bibinfo{person}{JoonHo Lee} {and} \bibinfo{person}{Gyemin Lee}.} \bibinfo{year}{2023}\natexlab{}.
\newblock \showarticletitle{Feature Alignment by Uncertainty and Self-Training for Source-Free Unsupervised Domain Adaptation}.
\newblock \bibinfo{journal}{\emph{Neural Networks}}  \bibinfo{volume}{161} (\bibinfo{year}{2023}), \bibinfo{pages}{682--692}.
\newblock
\showISSN{0893-6080}
\href{https://doi.org/10.1016/j.neunet.2023.02.009}{doi:\nolinkurl{10.1016/j.neunet.2023.02.009}}


\bibitem[Li et~al\mbox{.}(2023)]%
        {li2023blip}
\bibfield{author}{\bibinfo{person}{Junnan Li}, \bibinfo{person}{Dongxu Li}, \bibinfo{person}{Silvio Savarese}, {and} \bibinfo{person}{Steven Hoi}.} \bibinfo{year}{2023}\natexlab{}.
\newblock \showarticletitle{Blip-2: Bootstrapping language-image pre-training with frozen image encoders and large language models}. In \bibinfo{booktitle}{\emph{International conference on machine learning}}. PMLR, \bibinfo{pages}{19730--19742}.
\newblock


\bibitem[Li et~al\mbox{.}(2022b)]%
        {li2022blip}
\bibfield{author}{\bibinfo{person}{Junnan Li}, \bibinfo{person}{Dongxu Li}, \bibinfo{person}{Caiming Xiong}, {and} \bibinfo{person}{Steven Hoi}.} \bibinfo{year}{2022}\natexlab{b}.
\newblock \showarticletitle{Blip: Bootstrapping language-image pre-training for unified vision-language understanding and generation}. In \bibinfo{booktitle}{\emph{International conference on machine learning}}. PMLR, \bibinfo{pages}{12888--12900}.
\newblock


\bibitem[Li et~al\mbox{.}(2021)]%
        {li2021align}
\bibfield{author}{\bibinfo{person}{Junnan Li}, \bibinfo{person}{Ramprasaath Selvaraju}, \bibinfo{person}{Akhilesh Gotmare}, \bibinfo{person}{Shafiq Joty}, \bibinfo{person}{Caiming Xiong}, {and} \bibinfo{person}{Steven Chu~Hong Hoi}.} \bibinfo{year}{2021}\natexlab{}.
\newblock \showarticletitle{Align before fuse: Vision and language representation learning with momentum distillation}.
\newblock \bibinfo{journal}{\emph{Advances in neural information processing systems}}  \bibinfo{volume}{34} (\bibinfo{year}{2021}), \bibinfo{pages}{9694--9705}.
\newblock


\bibitem[Li et~al\mbox{.}(2022a)]%
        {li2022global}
\bibfield{author}{\bibinfo{person}{Yao Li}, \bibinfo{person}{Zhuoyuan Li}, \bibinfo{person}{Li Li}, \bibinfo{person}{Dong Liu}, {and} \bibinfo{person}{Houqiang Li}.} \bibinfo{year}{2022}\natexlab{a}.
\newblock \showarticletitle{Global homography motion compensation for versatile video coding}. In \bibinfo{booktitle}{\emph{2022 IEEE International Conference on Visual Communications and Image Processing (VCIP)}}. IEEE, \bibinfo{pages}{1--5}.
\newblock


\bibitem[Li et~al\mbox{.}(2024a)]%
        {li2024inloop}
\bibfield{author}{\bibinfo{person}{Zhuoyuan Li}, \bibinfo{person}{Jiacheng Li}, \bibinfo{person}{Yao Li}, \bibinfo{person}{Li Li}, \bibinfo{person}{Dong Liu}, {and} \bibinfo{person}{Feng Wu}.} \bibinfo{year}{2024}\natexlab{a}.
\newblock \showarticletitle{In-Loop Filtering via Trained Look-Up Tables}. In \bibinfo{booktitle}{\emph{2024 IEEE International Conference on Visual Communications and Image Processing (VCIP)}}. \bibinfo{pages}{1--5}.
\newblock


\bibitem[Li et~al\mbox{.}(2024b)]%
        {li2024uniformly}
\bibfield{author}{\bibinfo{person}{Zhuoyuan Li}, \bibinfo{person}{Yao Li}, \bibinfo{person}{Chuanbo Tang}, \bibinfo{person}{Li Li}, \bibinfo{person}{Dong Liu}, {and} \bibinfo{person}{Feng Wu}.} \bibinfo{year}{2024}\natexlab{b}.
\newblock \showarticletitle{Uniformly Accelerated Motion Model for Inter Prediction}. In \bibinfo{booktitle}{\emph{2024 IEEE International Conference on Visual Communications and Image Processing (VCIP)}}. \bibinfo{pages}{1--5}.
\newblock


\bibitem[Li et~al\mbox{.}(2024c)]%
        {li2024ustctd}
\bibfield{author}{\bibinfo{person}{Zhuoyuan Li}, \bibinfo{person}{Junqi Liao}, \bibinfo{person}{Chuanbo Tang}, \bibinfo{person}{Haotian Zhang}, \bibinfo{person}{Yuqi Li}, \bibinfo{person}{Yifan Bian}, \bibinfo{person}{Xihua Sheng}, \bibinfo{person}{Xinmin Feng}, \bibinfo{person}{Yao Li}, \bibinfo{person}{Changsheng Gao}, {et~al\mbox{.}}} \bibinfo{year}{2024}\natexlab{c}.
\newblock \showarticletitle{{USTC-TD}: A test dataset and benchmark for image and video coding in 2020s}.
\newblock \bibinfo{journal}{\emph{arXiv preprint arXiv:2409.08481}} (\bibinfo{year}{2024}).
\newblock


\bibitem[Li et~al\mbox{.}(2024d)]%
        {li2024object}
\bibfield{author}{\bibinfo{person}{Zhuoyuan Li}, \bibinfo{person}{Zikun Yuan}, \bibinfo{person}{Li Li}, \bibinfo{person}{Dong Liu}, \bibinfo{person}{Xiaohu Tang}, {and} \bibinfo{person}{Feng Wu}.} \bibinfo{year}{2024}\natexlab{d}.
\newblock \showarticletitle{Object Segmentation-Assisted Inter Prediction for Versatile Video Coding}.
\newblock \bibinfo{journal}{\emph{IEEE Transactions on Broadcasting}} \bibinfo{volume}{70}, \bibinfo{number}{4} (\bibinfo{year}{2024}), \bibinfo{pages}{1236--1253}.
\newblock


\bibitem[Liu et~al\mbox{.}(2024)]%
        {liu2024rate}
\bibfield{author}{\bibinfo{person}{Jinming Liu}, \bibinfo{person}{Ruoyu Feng}, \bibinfo{person}{Yunpeng Qi}, \bibinfo{person}{Qiuyu Chen}, \bibinfo{person}{Zhibo Chen}, \bibinfo{person}{Wenjun Zeng}, {and} \bibinfo{person}{Xin Jin}.} \bibinfo{year}{2024}\natexlab{}.
\newblock \showarticletitle{Rate-distortion-cognition controllable versatile neural image compression}. In \bibinfo{booktitle}{\emph{European Conference on Computer Vision}}. Springer, \bibinfo{pages}{329--348}.
\newblock


\bibitem[Liu et~al\mbox{.}(2021)]%
        {liu2021semantics}
\bibfield{author}{\bibinfo{person}{Kang Liu}, \bibinfo{person}{Dong Liu}, \bibinfo{person}{Li Li}, \bibinfo{person}{Ning Yan}, {and} \bibinfo{person}{Houqiang Li}.} \bibinfo{year}{2021}\natexlab{}.
\newblock \showarticletitle{Semantics-to-Signal Scalable Image Compression with Learned Revertible Representations}.
\newblock \bibinfo{journal}{\emph{Int. J. Comput. Vision}} \bibinfo{volume}{129}, \bibinfo{number}{9} (\bibinfo{date}{Sept.} \bibinfo{year}{2021}), \bibinfo{pages}{2605–2621}.
\newblock
\showISSN{0920-5691}
\href{https://doi.org/10.1007/s11263-021-01491-7}{doi:\nolinkurl{10.1007/s11263-021-01491-7}}


\bibitem[Long et~al\mbox{.}(2015)]%
        {long2015learning}
\bibfield{author}{\bibinfo{person}{Mingsheng Long}, \bibinfo{person}{Yue Cao}, \bibinfo{person}{Jianmin Wang}, {and} \bibinfo{person}{Michael Jordan}.} \bibinfo{year}{2015}\natexlab{}.
\newblock \showarticletitle{Learning transferable features with deep adaptation networks}. In \bibinfo{booktitle}{\emph{International conference on machine learning}}. PMLR, \bibinfo{pages}{97--105}.
\newblock


\bibitem[Ma et~al\mbox{.}(2024)]%
        {ma2024feature}
\bibfield{author}{\bibinfo{person}{Yifan Ma}, \bibinfo{person}{Changsheng Gao}, \bibinfo{person}{Qiaoxi Chen}, \bibinfo{person}{Li Li}, \bibinfo{person}{Dong Liu}, {and} \bibinfo{person}{Xiaoyan Sun}.} \bibinfo{year}{2024}\natexlab{}.
\newblock \showarticletitle{Feature Compression With 3D Sparse Convolution}. In \bibinfo{booktitle}{\emph{VCIP}}. \bibinfo{pages}{1--5}.
\newblock


\bibitem[Mao et~al\mbox{.}(2024)]%
        {mao2024perceptual}
\bibfield{author}{\bibinfo{person}{Rui Mao}, \bibinfo{person}{Xinmin Feng}, \bibinfo{person}{Changsheng Gao}, \bibinfo{person}{Li Li}, \bibinfo{person}{Dong Liu}, {and} \bibinfo{person}{Xiaoyan Sun}.} \bibinfo{year}{2024}\natexlab{}.
\newblock \showarticletitle{Perceptual Image Compression With Conditional Diffusion Transformers}. In \bibinfo{booktitle}{\emph{VCIP}}. \bibinfo{pages}{1--5}.
\newblock


\bibitem[Misra et~al\mbox{.}(2022)]%
        {misra2022video}
\bibfield{author}{\bibinfo{person}{Kiran Misra}, \bibinfo{person}{Tianying Ji}, \bibinfo{person}{Andrew Segall}, {and} \bibinfo{person}{Frank Bossen}.} \bibinfo{year}{2022}\natexlab{}.
\newblock \showarticletitle{Video Feature Compression for Machine Tasks}. In \bibinfo{booktitle}{\emph{ICME}}. \bibinfo{pages}{1--6}.
\newblock
\href{https://doi.org/10.1109/ICME52920.2022.9859894}{doi:\nolinkurl{10.1109/ICME52920.2022.9859894}}


\bibitem[Oquab et~al\mbox{.}(2023)]%
        {oquab2023dinov2}
\bibfield{author}{\bibinfo{person}{Maxime Oquab}, \bibinfo{person}{Timoth{\'e}e Darcet}, \bibinfo{person}{Th{\'e}o Moutakanni}, \bibinfo{person}{Huy Vo}, \bibinfo{person}{Marc Szafraniec}, \bibinfo{person}{Vasil Khalidov}, \bibinfo{person}{Pierre Fernandez}, \bibinfo{person}{Daniel Haziza}, \bibinfo{person}{Francisco Massa}, \bibinfo{person}{Alaaeldin El-Nouby}, {et~al\mbox{.}}} \bibinfo{year}{2023}\natexlab{}.
\newblock \showarticletitle{{DINOv2}: Learning robust visual features without supervision}.
\newblock \bibinfo{journal}{\emph{arXiv preprint arXiv:2304.07193}} (\bibinfo{year}{2023}).
\newblock


\bibitem[Radford et~al\mbox{.}(2021)]%
        {radford2021learning}
\bibfield{author}{\bibinfo{person}{Alec Radford}, \bibinfo{person}{Jong~Wook Kim}, \bibinfo{person}{Chris Hallacy}, \bibinfo{person}{Aditya Ramesh}, \bibinfo{person}{Gabriel Goh}, \bibinfo{person}{Sandhini Agarwal}, \bibinfo{person}{Girish Sastry}, \bibinfo{person}{Amanda Askell}, \bibinfo{person}{Pamela Mishkin}, \bibinfo{person}{Jack Clark}, {et~al\mbox{.}}} \bibinfo{year}{2021}\natexlab{}.
\newblock \showarticletitle{Learning transferable visual models from natural language supervision}. In \bibinfo{booktitle}{\emph{ICML}}. PMLR, \bibinfo{pages}{8748--8763}.
\newblock


\bibitem[Sagawa* et~al\mbox{.}(2020)]%
        {Sagawa2020Distributionally}
\bibfield{author}{\bibinfo{person}{Shiori Sagawa*}, \bibinfo{person}{Pang~Wei Koh*}, \bibinfo{person}{Tatsunori~B. Hashimoto}, {and} \bibinfo{person}{Percy Liang}.} \bibinfo{year}{2020}\natexlab{}.
\newblock \showarticletitle{Distributionally Robust Neural Networks}. In \bibinfo{booktitle}{\emph{ICLR}}.
\newblock


\bibitem[Senushkin et~al\mbox{.}(2023)]%
        {senushkin2023independent}
\bibfield{author}{\bibinfo{person}{Dmitry Senushkin}, \bibinfo{person}{Nikolay Patakin}, \bibinfo{person}{Arseny Kuznetsov}, {and} \bibinfo{person}{Anton Konushin}.} \bibinfo{year}{2023}\natexlab{}.
\newblock \showarticletitle{Independent component alignment for multi-task learning}. In \bibinfo{booktitle}{\emph{Proceedings of the IEEE/CVF Conference on Computer Vision and Pattern Recognition}}. \bibinfo{pages}{20083--20093}.
\newblock


\bibitem[Shen et~al\mbox{.}(2024a)]%
        {shen2024image}
\bibfield{author}{\bibinfo{person}{Xuelin Shen}, \bibinfo{person}{Haoqiao Ou}, {and} \bibinfo{person}{Wenhan Yang}.} \bibinfo{year}{2024}\natexlab{a}.
\newblock \showarticletitle{Image Coding For Machine Via Analytics-Driven Appearance Redundancy Reduction}. In \bibinfo{booktitle}{\emph{2024 IEEE International Conference on Image Processing (ICIP)}}. IEEE, \bibinfo{pages}{1883--1889}.
\newblock


\bibitem[Shen et~al\mbox{.}(2024b)]%
        {shen2024image2}
\bibfield{author}{\bibinfo{person}{Xuelin Shen}, \bibinfo{person}{Kangsheng Yin}, \bibinfo{person}{Xu Wang}, \bibinfo{person}{Yulin He}, \bibinfo{person}{Shiqi Wang}, {and} \bibinfo{person}{Wenhan Yang}.} \bibinfo{year}{2024}\natexlab{b}.
\newblock \showarticletitle{Image coding for analytics via adversarially augmented adaptation}. In \bibinfo{booktitle}{\emph{ICASSP 2024-2024 IEEE International Conference on Acoustics, Speech and Signal Processing (ICASSP)}}. IEEE, \bibinfo{pages}{3605--3609}.
\newblock


\bibitem[Sun and Saenko(2016)]%
        {sun2016deep}
\bibfield{author}{\bibinfo{person}{Baochen Sun} {and} \bibinfo{person}{Kate Saenko}.} \bibinfo{year}{2016}\natexlab{}.
\newblock \showarticletitle{Deep coral: Correlation alignment for deep domain adaptation}. In \bibinfo{booktitle}{\emph{Computer vision--ECCV 2016 workshops: Amsterdam, the Netherlands, October 8-10 and 15-16, 2016, proceedings, part III 14}}. Springer, \bibinfo{pages}{443--450}.
\newblock


\bibitem[Tang et~al\mbox{.}(2024)]%
        {tang2024offline}
\bibfield{author}{\bibinfo{person}{Chuanbo Tang}, \bibinfo{person}{Xihua Sheng}, \bibinfo{person}{Zhuoyuan Li}, \bibinfo{person}{Haotian Zhang}, \bibinfo{person}{Li Li}, {and} \bibinfo{person}{Dong Liu}.} \bibinfo{year}{2024}\natexlab{}.
\newblock \showarticletitle{Offline and online optical flow enhancement for deep video compression}. In \bibinfo{booktitle}{\emph{Proceedings of the AAAI Conference on Artificial Intelligence}}, Vol.~\bibinfo{volume}{38}. \bibinfo{pages}{5118--5126}.
\newblock


\bibitem[Team et~al\mbox{.}(2024)]%
        {team2024gemini}
\bibfield{author}{\bibinfo{person}{Gemini Team}, \bibinfo{person}{Petko Georgiev}, \bibinfo{person}{Ving~Ian Lei}, \bibinfo{person}{Ryan Burnell}, \bibinfo{person}{Libin Bai}, \bibinfo{person}{Anmol Gulati}, \bibinfo{person}{Garrett Tanzer}, \bibinfo{person}{Damien Vincent}, \bibinfo{person}{Zhufeng Pan}, \bibinfo{person}{Shibo Wang}, {et~al\mbox{.}}} \bibinfo{year}{2024}\natexlab{}.
\newblock \showarticletitle{Gemini 1.5: Unlocking multimodal understanding across millions of tokens of context}.
\newblock \bibinfo{journal}{\emph{arXiv preprint arXiv:2403.05530}} (\bibinfo{year}{2024}).
\newblock


\bibitem[Tian et~al\mbox{.}(2024b)]%
        {tian2024coding}
\bibfield{author}{\bibinfo{person}{Yuan Tian}, \bibinfo{person}{Guo Lu}, \bibinfo{person}{Yichao Yan}, \bibinfo{person}{Guangtao Zhai}, \bibinfo{person}{Li Chen}, {and} \bibinfo{person}{Zhiyong Gao}.} \bibinfo{year}{2024}\natexlab{b}.
\newblock \showarticletitle{A coding framework and benchmark towards low-bitrate video understanding}.
\newblock \bibinfo{journal}{\emph{IEEE Transactions on Pattern Analysis and Machine Intelligence}} \bibinfo{volume}{46}, \bibinfo{number}{8} (\bibinfo{year}{2024}), \bibinfo{pages}{5852--5872}.
\newblock


\bibitem[Tian et~al\mbox{.}(2024a)]%
        {tian2024free}
\bibfield{author}{\bibinfo{person}{Yuan Tian}, \bibinfo{person}{Guo Lu}, {and} \bibinfo{person}{Guangtao Zhai}.} \bibinfo{year}{2024}\natexlab{a}.
\newblock \showarticletitle{Free-VSC: Free semantics from visual foundation models for unsupervised video semantic compression}. In \bibinfo{booktitle}{\emph{European Conference on Computer Vision}}. Springer, \bibinfo{pages}{163--183}.
\newblock


\bibitem[Tian et~al\mbox{.}(2023)]%
        {tian2023non}
\bibfield{author}{\bibinfo{person}{Yuan Tian}, \bibinfo{person}{Guo Lu}, \bibinfo{person}{Guangtao Zhai}, {and} \bibinfo{person}{Zhiyong Gao}.} \bibinfo{year}{2023}\natexlab{}.
\newblock \showarticletitle{Non-semantics suppressed mask learning for unsupervised video semantic compression}. In \bibinfo{booktitle}{\emph{Proceedings of the IEEE/CVF International Conference on Computer Vision}}. \bibinfo{pages}{13610--13622}.
\newblock


\bibitem[Tian et~al\mbox{.}(2022)]%
        {tian2022fedbert}
\bibfield{author}{\bibinfo{person}{Yuanyishu Tian}, \bibinfo{person}{Yao Wan}, \bibinfo{person}{Lingjuan Lyu}, \bibinfo{person}{Dezhong Yao}, \bibinfo{person}{Hai Jin}, {and} \bibinfo{person}{Lichao Sun}.} \bibinfo{year}{2022}\natexlab{}.
\newblock \showarticletitle{{FedBERT:} When federated learning meets pre-training}.
\newblock \bibinfo{journal}{\emph{ACM Transactions on Intelligent Systems and Technology}} \bibinfo{volume}{13}, \bibinfo{number}{4} (\bibinfo{year}{2022}), \bibinfo{pages}{1--26}.
\newblock


\bibitem[Touvron et~al\mbox{.}(2023)]%
        {touvron2023llama}
\bibfield{author}{\bibinfo{person}{Hugo Touvron}, \bibinfo{person}{Thibaut Lavril}, \bibinfo{person}{Gautier Izacard}, \bibinfo{person}{Xavier Martinet}, \bibinfo{person}{Marie-Anne Lachaux}, \bibinfo{person}{Timoth{\'e}e Lacroix}, \bibinfo{person}{Baptiste Rozi{\`e}re}, \bibinfo{person}{Naman Goyal}, \bibinfo{person}{Eric Hambro}, \bibinfo{person}{Faisal Azhar}, {et~al\mbox{.}}} \bibinfo{year}{2023}\natexlab{}.
\newblock \showarticletitle{Llama: Open and efficient foundation language models}.
\newblock \bibinfo{journal}{\emph{arXiv preprint arXiv:2302.13971}} (\bibinfo{year}{2023}).
\newblock


\bibitem[Wang et~al\mbox{.}(2022b)]%
        {wang2022towards}
\bibfield{author}{\bibinfo{person}{Shurun Wang}, \bibinfo{person}{Shiqi Wang}, \bibinfo{person}{Wenhan Yang}, \bibinfo{person}{Xinfeng Zhang}, \bibinfo{person}{Shanshe Wang}, \bibinfo{person}{Siwei Ma}, {and} \bibinfo{person}{Wen Gao}.} \bibinfo{year}{2022}\natexlab{b}.
\newblock \showarticletitle{Towards Analysis-Friendly Face Representation With Scalable Feature and Texture Compression}.
\newblock \bibinfo{journal}{\emph{IEEE Transactions on Multimedia}}  \bibinfo{volume}{24} (\bibinfo{year}{2022}), \bibinfo{pages}{3169--3181}.
\newblock
\href{https://doi.org/10.1109/TMM.2021.3094300}{doi:\nolinkurl{10.1109/TMM.2021.3094300}}


\bibitem[Wang et~al\mbox{.}(2023b)]%
        {wang2023feature}
\bibfield{author}{\bibinfo{person}{Shuai Wang}, \bibinfo{person}{Daoan Zhang}, \bibinfo{person}{Zipei Yan}, \bibinfo{person}{Jianguo Zhang}, {and} \bibinfo{person}{Rui Li}.} \bibinfo{year}{2023}\natexlab{b}.
\newblock \showarticletitle{Feature alignment and uniformity for test time adaptation}. In \bibinfo{booktitle}{\emph{Proceedings of the IEEE/CVF Conference on Computer Vision and Pattern Recognition}}. \bibinfo{pages}{20050--20060}.
\newblock


\bibitem[Wang et~al\mbox{.}(2023a)]%
        {wang2023intermediate}
\bibfield{author}{\bibinfo{person}{Weiqian Wang}, \bibinfo{person}{Ping An}, \bibinfo{person}{Xinpeng Huang}, \bibinfo{person}{Kunqiang Huang}, {and} \bibinfo{person}{Chao Yang}.} \bibinfo{year}{2023}\natexlab{a}.
\newblock \showarticletitle{Intermediate deep feature coding for human--machine vision collaboration}.
\newblock \bibinfo{journal}{\emph{Journal of Visual Communication and Image Representation}}  \bibinfo{volume}{95} (\bibinfo{year}{2023}), \bibinfo{pages}{103859}.
\newblock


\bibitem[Wang et~al\mbox{.}(2022a)]%
        {wang2022domain}
\bibfield{author}{\bibinfo{person}{Xiaoling Wang}, \bibinfo{person}{Qi Kang}, \bibinfo{person}{MengChu Zhou}, \bibinfo{person}{Siya Yao}, {and} \bibinfo{person}{Abdullah Abusorrah}.} \bibinfo{year}{2022}\natexlab{a}.
\newblock \showarticletitle{Domain adaptation multitask optimization}.
\newblock \bibinfo{journal}{\emph{IEEE Transactions on Cybernetics}} \bibinfo{volume}{53}, \bibinfo{number}{7} (\bibinfo{year}{2022}), \bibinfo{pages}{4567--4578}.
\newblock


\bibitem[Wang et~al\mbox{.}(2024)]%
        {wang2024low}
\bibfield{author}{\bibinfo{person}{Zixi Wang}, \bibinfo{person}{Fan Li}, \bibinfo{person}{Yunfei Zhang}, {and} \bibinfo{person}{Yuan Zhang}.} \bibinfo{year}{2024}\natexlab{}.
\newblock \showarticletitle{Low-Rate Feature Compression for Collaborative Intelligence: Reducing Redundancy in Spatial and Statistical Levels}.
\newblock \bibinfo{journal}{\emph{IEEE Transactions on Multimedia}}  \bibinfo{volume}{26} (\bibinfo{year}{2024}), \bibinfo{pages}{2756--2771}.
\newblock
\href{https://doi.org/10.1109/TMM.2023.3303716}{doi:\nolinkurl{10.1109/TMM.2023.3303716}}


\bibitem[WG2(2023)]%
        {fcm_cfp}
\bibfield{author}{\bibinfo{person}{WG2}.} \bibinfo{year}{April 2023}\natexlab{}.
\newblock \showarticletitle{Call for Proposals on Feature Compression for Video Coding for Machines}.
\newblock  \bibinfo{volume}{ISO/IEC JTC 1/SC 29/WG 2}, \bibinfo{number}{N282} (\bibinfo{year}{April 2023}).
\newblock


\bibitem[Wu et~al\mbox{.}(2025)]%
        {wu2025codebook}
\bibfield{author}{\bibinfo{person}{Xu Wu}, \bibinfo{person}{Xianxu Hou}, \bibinfo{person}{Zhihui Lai}, \bibinfo{person}{Jie Zhou}, \bibinfo{person}{Ya-nan Zhang}, \bibinfo{person}{Witold Pedrycz}, {and} \bibinfo{person}{Linlin Shen}.} \bibinfo{year}{2025}\natexlab{}.
\newblock \showarticletitle{A codebook-driven approach for low-light image enhancement}.
\newblock \bibinfo{journal}{\emph{Engineering Applications of Artificial Intelligence}}  \bibinfo{volume}{156} (\bibinfo{year}{2025}), \bibinfo{pages}{111115}.
\newblock


\bibitem[Yan et~al\mbox{.}(2021)]%
        {yan2021SSSIC}
\bibfield{author}{\bibinfo{person}{Ning Yan}, \bibinfo{person}{Changsheng Gao}, \bibinfo{person}{Dong Liu}, \bibinfo{person}{Houqiang Li}, \bibinfo{person}{Li Li}, {and} \bibinfo{person}{Feng Wu}.} \bibinfo{year}{2021}\natexlab{}.
\newblock \showarticletitle{{SSSIC}: Semantics-to-Signal Scalable Image Coding With Learned Structural Representations}.
\newblock \bibinfo{journal}{\emph{IEEE Transactions on Image Processing}}  \bibinfo{volume}{30} (\bibinfo{year}{2021}), \bibinfo{pages}{8939--8954}.
\newblock
\href{https://doi.org/10.1109/TIP.2021.3121131}{doi:\nolinkurl{10.1109/TIP.2021.3121131}}


\bibitem[Ye et~al\mbox{.}(2024)]%
        {ye2024openfedllm}
\bibfield{author}{\bibinfo{person}{Rui Ye}, \bibinfo{person}{Wenhao Wang}, \bibinfo{person}{Jingyi Chai}, \bibinfo{person}{Dihan Li}, \bibinfo{person}{Zexi Li}, \bibinfo{person}{Yinda Xu}, \bibinfo{person}{Yaxin Du}, \bibinfo{person}{Yanfeng Wang}, {and} \bibinfo{person}{Siheng Chen}.} \bibinfo{year}{2024}\natexlab{}.
\newblock \showarticletitle{{OpenFedLLM}: Training large language models on decentralized private data via federated learning}. In \bibinfo{booktitle}{\emph{ACM SIGKDD}}. \bibinfo{pages}{6137--6147}.
\newblock


\bibitem[Yeh et~al\mbox{.}(2021)]%
        {yeh2021sofa}
\bibfield{author}{\bibinfo{person}{Hao-Wei Yeh}, \bibinfo{person}{Baoyao Yang}, \bibinfo{person}{Pong~C Yuen}, {and} \bibinfo{person}{Tatsuya Harada}.} \bibinfo{year}{2021}\natexlab{}.
\newblock \showarticletitle{{SOFA}: Source-data-free feature alignment for unsupervised domain adaptation}. In \bibinfo{booktitle}{\emph{Proceedings of the IEEE/CVF Winter Conference on Applications of Computer Vision}}. \bibinfo{pages}{474--483}.
\newblock


\bibitem[Yin et~al\mbox{.}(2025)]%
        {yin2025unified}
\bibfield{author}{\bibinfo{person}{Kangsheng Yin}, \bibinfo{person}{Quan Liu}, \bibinfo{person}{Xuelin Shen}, \bibinfo{person}{Yulin He}, \bibinfo{person}{Wenhan Yang}, {and} \bibinfo{person}{Shiqi Wang}.} \bibinfo{year}{2025}\natexlab{}.
\newblock \showarticletitle{Unified Coding for Both Human Perception and Generalized Machine Analytics with CLIP Supervision}. In \bibinfo{booktitle}{\emph{Proceedings of the AAAI Conference on Artificial Intelligence}}, Vol.~\bibinfo{volume}{39}. \bibinfo{pages}{9517--9525}.
\newblock


\bibitem[Zhang et~al\mbox{.}(2021a)]%
        {zhang2021understanding}
\bibfield{author}{\bibinfo{person}{Chiyuan Zhang}, \bibinfo{person}{Samy Bengio}, \bibinfo{person}{Moritz Hardt}, \bibinfo{person}{Benjamin Recht}, {and} \bibinfo{person}{Oriol Vinyals}.} \bibinfo{year}{2021}\natexlab{a}.
\newblock \showarticletitle{Understanding deep learning (still) requires rethinking generalization}.
\newblock \bibinfo{journal}{\emph{Commun. ACM}} \bibinfo{volume}{64}, \bibinfo{number}{3} (\bibinfo{date}{Feb.} \bibinfo{year}{2021}), \bibinfo{pages}{107–115}.
\newblock
\showISSN{0001-0782}
\href{https://doi.org/10.1145/3446776}{doi:\nolinkurl{10.1145/3446776}}


\bibitem[Zhang et~al\mbox{.}(2021b)]%
        {zhang2021MSFC}
\bibfield{author}{\bibinfo{person}{Zhicong Zhang}, \bibinfo{person}{Mengyang Wang}, \bibinfo{person}{Mengyao Ma}, \bibinfo{person}{Jiahui Li}, {and} \bibinfo{person}{Xiaopeng Fan}.} \bibinfo{year}{2021}\natexlab{b}.
\newblock \showarticletitle{{MSFC}: Deep Feature Compression in Multi-Task Network}. In \bibinfo{booktitle}{\emph{ICME}}. \bibinfo{pages}{1--6}.
\newblock
\href{https://doi.org/10.1109/ICME51207.2021.9428258}{doi:\nolinkurl{10.1109/ICME51207.2021.9428258}}


\bibitem[Zhu et~al\mbox{.}(2024)]%
        {zhu2024learned}
\bibfield{author}{\bibinfo{person}{Lingyu Zhu}, \bibinfo{person}{Binzhe Li}, \bibinfo{person}{Riyu Lu}, \bibinfo{person}{Peilin Chen}, \bibinfo{person}{Qi Mao}, \bibinfo{person}{Zhao Wang}, \bibinfo{person}{Wenhan Yang}, {and} \bibinfo{person}{Shiqi Wang}.} \bibinfo{year}{2024}\natexlab{}.
\newblock \showarticletitle{Learned Image Compression for Both Humans and Machines via Dynamic Adaptation}. In \bibinfo{booktitle}{\emph{2024 IEEE International Conference on Image Processing (ICIP)}}. \bibinfo{pages}{1788--1794}.
\newblock


\bibitem[Zuo et~al\mbox{.}(2024)]%
        {zuo2024learned}
\bibfield{author}{\bibinfo{person}{Yanchen Zuo}, \bibinfo{person}{Changsheng Gao}, \bibinfo{person}{Dong Liu}, \bibinfo{person}{Li Li}, \bibinfo{person}{Yueyi Zhang}, {and} \bibinfo{person}{Xiaoyan Sun}.} \bibinfo{year}{2024}\natexlab{}.
\newblock \showarticletitle{Learned Rate-Distortion Cost Prediction for Ultrafast Screen Content Intra Coding}.
\newblock \bibinfo{journal}{\emph{IEEE Transactions on Circuits and Systems for Video Technology}} \bibinfo{volume}{34}, \bibinfo{number}{3} (\bibinfo{year}{2024}), \bibinfo{pages}{1976--1980}.
\newblock


\end{thebibliography}

\end{document}